\documentclass[twocol]{article}

\usepackage{url}

\usepackage{soul}

\usepackage{geometry}

\usepackage{longtable}

\usepackage{algorithmic}
\usepackage{etoolbox}
\usepackage{tikz}

\usepackage{bm}
\usepackage{blkarray}
\usepackage{cancel}
\usepackage{multirow}
\usepackage{booktabs}
\usepackage{algorithm2e}
\usepackage{natbib}
\usepackage[font=small,width=0.88\textwidth]{caption}

\usepackage{amsmath,amsfonts,amssymb,bm, amsthm}
\newcommand\new[1]{{#1}}
\usepackage{cleveref}
\newcommand\given[1][]{\:#1\vert\:}

\DeclareMathOperator*{\argmax}{arg\,max}
\DeclareMathOperator*{\argmin}{arg\,min}

\DeclareMathOperator{\Ex}{\mathbb{E}}
\DeclareMathOperator{\Var}{\text{Var}}
\newcommand{\mlpath}{\hat{\mathbf{p}}}

\newtheorem{mydef}{Definition}

\begin{document}

	  \author{Michael O'Malley$^{1}$, Adam M. Sykulski$^{1}$,\\Romuald Laso-Jadart$^2$, Mohammed-Amin Madoui$^{2}$\let\thefootnote\relax\footnote{$^1$ STOR-i Centre for Doctoral Training~/~Department of Mathematics and Statistics, Lancaster University, UK}\let\thefootnote\relax\footnote{$^2$ Génomique Métabolique, Genoscope, Institut F. Jacob, CEA, CNRS, Univ Evry, Univ Paris-Saclay, Evry, France}}

		\title{Estimating the travel time and the most likely path from Lagrangian drifters}
		\date{}
\maketitle
	\begin{abstract}
		We provide a novel methodology for computing the most likely path taken by drifters between arbitrary fixed locations in the ocean. We also provide an estimate of the travel time associated with this path. Lagrangian pathways and travel times are of practical value not just in understanding surface velocities, but also in modelling the transport of ocean-borne species such as planktonic organisms, and floating debris such as plastics. In particular, the estimated travel time can be used to compute an estimated Lagrangian distance, which is often more informative than Euclidean distance in understanding connectivity between locations. Our methodology is purely data-driven, and requires no simulations of drifter trajectories, in contrast to existing approaches. Our method scales globally and can simultaneously handle multiple locations in the ocean. Furthermore, we provide estimates of the error and uncertainty associated with both the most likely path and the associated travel time.
	\end{abstract}
	
	\section{Introduction}
	The Lagrangian study of transport and mixing in the ocean is of fundamental interest to ocean modellers \citep{VANSEBILLE2018, VanSebille2009, LaCasce2008}. In particular, the analysis of data obtained from Lagrangian drifting objects greatly contribute to our knowledge of ocean circulation, e.g. through analysing the accuracy of numerical and stochastic models \citep{huntley2011lagrangian,Sykulski2016}, or the use of drifter data to better understand various pathways and where to search for marine debris \citep{miron2019markov, VanSebille2012,mcadam2018surface}.
	
	\citet{meehl1982characteristics} used shipdrift data to create a surface velocity data set on a $5^\circ\times 5^\circ$ grid. These velocities were used to simulate the Lagrangian drift of floating objects in \citet{wakata1990lagrangian}. More recent works focus on using drifting buoys to derive Lagrangian models to discover areas where floating debris tends to end up \citep{VanSebille2014, VanSebille2012, Maximenko2012}. Advances in technology have resulted in much better data quality, which now permits the use of more detailed methodology. The newer models provide densities of where debris ends up on grid scales as small as $0.5^\circ \times 0.5^\circ$.
	
	In this paper, we propose a novel computationally fast method for estimating a so called \textit{``most likely pathway"} between two points in the ocean, alongside an estimated travel time for this pathway. The method is purely data-driven. We demonstrate our methodology on data from the \textit{Global Drifter Program (GDP)}, but the method is designed to work with any Lagrangian tracking data set. Additionally, we develop and test related methodology for providing uncertainty on both the pathways and the travel times. Our method is automated with little expert knowledge needed from the practitioner. We provide a set of default parameters which allow the method to run as intended. The method simply takes in a set of locations within the ocean, and outputs a data structure containing most likely paths and corresponding travel time estimates between all pairs of locations. We focus on a global scale: we aim to provide a measure of Lagrangian connectivity for locations which are thousands of kilometres apart. An individual drifter trajectory is unlikely to connect two arbitrary locations far apart, hence the need for our methodology which fuses information across many drifters.
	
	A tool which predicts travel times is of practical use in many fields. For example in ecological studies of marine species, genetic measurements are taken at various locations in the ocean \citep{watsonabstract}. Euclidean distance is often used as a measure of separability and isolation-by-distance \citep{becking2006beta, Ellingsen2002} to find correlations with diversity metrics or genetic differentiation between communities or populations of organisms. Due to various currents and land barriers, we expect Euclidean distance to often be a poor measure of how `distant' or dissimilar communities or populations sampled in two locations are. The method proposed in this work would use the estimated travel times to supply a matrix containing a {\em Lagrangian distance} measure between all pairs of locations. This matrix can then be contrasted with a pairwise genetic distance matrix between these locations and will yield new insights. In many instances the Lagrangian distance matrix will be more correlated with genetic relatedness than a Euclidean distance matrix. This observation was already made in the Mediterranean Sea when studying plankton~\citep{Berline2014}, and off the coast of California for a species of sea snail \citep{white2010ocean}. Both of the works by \citet{Berline2014} and \citet{white2010ocean} rely on simulating trajectories from detailed ocean current data sets to estimate the Lagrangian distance. Such approaches do not scale globally and rely on simulated trajectories from currents rather than real observations.
	
\begin{figure*}[ht]
		\includegraphics[width=\textwidth]{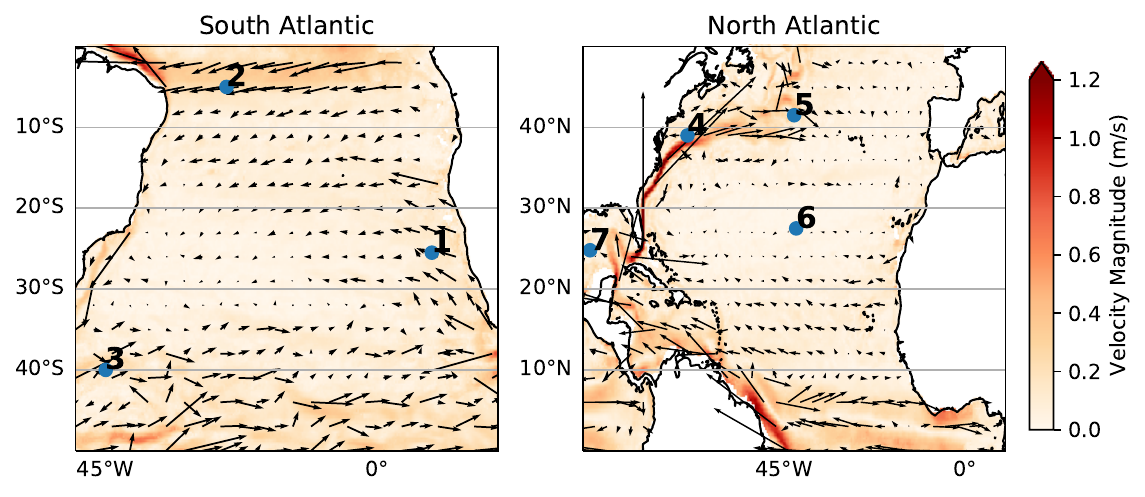}
		\caption{Locations of interest from Table \ref{tbl:locations}. Annual mean values of the near-surface currents derived from drifter velocities \citep{laurindo2017improved} are plotted. Arrows drawn on a $3^\circ\times 3^\circ$ grid to show current direction.}
		\label{fig:Station_loc}
	\end{figure*}

	In \Cref{fig:Station_loc}, we show seven locations plotted on a map with ocean currents. We use these locations as a proof-of-concept example throughout this paper. The exact coordinates are given in \Cref{tbl:locations}. The aim is to introduce and motivate a method which provides an estimate as to how long it would take to drift between any two of these locations. For example, the travel time from location 2 to location 3 in the South Atlantic Ocean should be smaller than the return journey due to the Brazil current. We choose to include locations in both the North and South Atlantic as we wish to demonstrate that the method successfully finds pathways linking points which are extremely far apart.
	
	\subsection{Comparison with Related Works}
	In this section we give a brief overview of techniques that have used the Global Drifter Program to achieve a similar or related task. The work by \citet{Rypina2017} proposes a statistical approach for obtaining travel times. A source area is defined such that at least 100 drifters pass through the source area. The method focuses on obtaining a spatial probability map and a mean travel time for every $1^\circ \times 1^\circ $ bin outside of the source area. This method successfully combines many trajectories, however for multiple locations one would have to decide on a varying grid box for each location of interest. Such a grid box must be manually chosen by the practitioner meaning that the method does not scale well with multiple locations. \citet{Rypina2017} also focus on estimating a mean travel time, where our method provides a travel time associated with the most likely path, and is hence more akin to estimating a mode or median travel time.

	The method by \citet{Sebille2011}, which proposes the use of Monte Carlo Super Trajectories (MCST), could naturally be used to estimate travel times. This method simulates new trajectories as sequences of unique grid indices along with corresponding travel time estimates for each part of that journey. The method is purely data driven i.e. they only use real trajectories to fit the model. The travel time and pathway we supply here should be similar to the most likely MCST to occur between the two points. The advantage of our methodology is that we do not base the analysis on a simulation, such that the results from the method described in \Cref{sec:method} are not subject to any randomness due to simulation.
	
	Various other works have made attempts to compute Lagrangian based distances. For example, \citet{Berline2014} used numerically simulated trajectories to estimate \textit{Mean Connection Times} within the Mediterranean Sea. \citet{Smith2018} used MCST to estimate various statistics of how seagrass fragments could drift from the South East coast of Australia to Chile. Specifically, \citet{Smith2018} simulated 10 million MCST starting from the SE coast of Australia and only 264 (0.00264\%) of the simulated trajectories were found to travel roughly to the Chilean coast.
	
	The approach by \citet{jonsson2016timescales} uses simulated drifter data to construct connectivity matrices between locations in the ocean. As the matrix is sparse, Dijkstra’s algorithm is used to connect arbitrarily distant locations in the ocean to measure Lagrangian distance. Although this method may at first glance bare similarities with our method (specifically in the use of Dijkstra’s algorithm), there are in fact many differences. First of all, the method uses simulated trajectories whereas we use real drifter trajectories. Secondly, Dijkstra’s algorithm is performed by \citet{jonsson2016timescales} on the {\em connectivity} matrix (which finds minimum connection times between locations), whereas our approach uses Dijkstra’s algorithm on the {\em transition} matrix which describes a probabilistic framework for drifter movement. We found the latter approach to perform much better with real data. Finally, we cannot directly implement the approach described in \citet{jonsson2016timescales} as only connectivity values higher than one year are \new{used by the algorithm}. For real data such a step would result in a very sparse connectivity matrix making the method infeasible. An initial analysis we conducted using similar methodology achieved poor results.
	
  There are a variety of works which use Markov transition matrices for different aims to this work. \cite{ser2015most} and \cite{miron2019markov} look at probable paths, where both of these works find a path going between two points in a certain number of days using a dynamic program. 
  \cite{froyland2014well} and \cite{miron2017lagrangian} study ocean dynamics by analysing eigenvalues of the transition matrix.  Other methods in the literature include characterizing dispersion and mixing \citep{ser2015flow}, identification coherent regions \citep{froyland2007detection,ser2015flow}, forward integration of tracers \citep{VanSebille2012, Maximenko2012}, and guiding drifter deployments \citep{lumpkin2016fulfilling}. We differ from these works as we ultimately aim to find travel times, as well as pathways, between multiple fixed locations.

    \new{Our proposed algorithm for computing travel times and pathways will also use the aforementioned Markov transition matrix approach. Our key novelty is that we build on this conceptual approach by implementing and demonstrating the benefits of using the (H3) spatial indexing system for discretization, and by supplying uncertainty quantification guidelines by applying grid rotations and data bootstrapping. The steps outlined in \Cref{alg:method} are individually known across disparate literature, however, this is the first paper to our knowledge that effectively combines these steps to solve the problem of interest. We provide numerous examples to show how our methodology robustly outperforms state-of-the-art alternative approaches. In addition, we supply freely-available software in the form of a Python package, of which all parameters in the model can easily be customized to suit the needs of the practitioner.}
    
    In summary, the novel contributions of this work are: \textit{a)} the combination of the steps in Section \ref{sec:method} to form a computationally-efficient algorithm which applies directly to transition matrices to find most likely paths and travel times simultaneously, \textit{b)} computation of uncertainty from discretization error and data sampling (Section~\ref{sec:stability}), and \textit{c)} the demonstration of the method showing it successfully obtains robust measures of connectivity between both very distant and closely located points (Section~\ref{sec:results}). The key outcome is that we obtain oceanographic travel times and most likely paths requiring no simulated trajectories.
    
	
	We believe our method is preferable to \cite{Rypina2017} as we do not require custom treatment to different source areas. \cite{jonsson2016timescales} requires the simulation of many very long and expensive to compute trajectories which obtain spurious results on real data. Using MCST's as in \cite{Smith2018} relies on simulation.
    The estimation of a full pairwise travel time matrix of the locations in Table \ref{tbl:locations} requires 42 travel time estimations. With MCSTs this would likely require the simulation of millions of trajectories and manual analysis of each location pair. Our method, in contrast, can produce such a travel time matrix in a matter of seconds given that the transition matrix needs to be estimated just once {\em a priori}. In a similar manner, global travel time maps can be made in a matter of minutes, such as those that we will be showing in Section \ref{sec:results}.


	\section{Background and Notation}
	\label{sec:background}
	\subsection{Global Drifter Program}
	The Global Drifter Program(GDP) is a database managed by the National Oceanographic and Atmospheric Administraction (NOAA) \citep{GDP, Lumpkin2007}. This data set contains over 20,000 free-floating buoys temporally spanning from February 15, 1979 through to the current day. These buoys are referred to as \textit{drifters}. The drifter design comprises of a sub-surface float and a drogue sock. Often this drogue sock detaches. We refer to the drifters which have lost their drogue sock as non-drogued drifters, and drogued for those which still have the drogue attached.
	
\begin{figure*}[ht]
		\centerline{
			\includegraphics[]{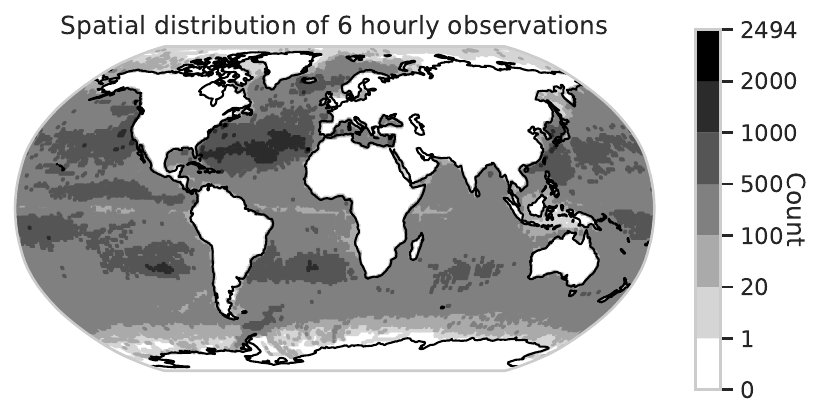}
		}
		\caption{Number of observations from the Global Drifter Program in each $1^\circ\times 1^\circ$ longitude-latitude box.}\label{fig:spatial_dist}
	\end{figure*}

	Here we use the drifter data recorded up to July, 2020. We use data which has been recorded from drogued drifters only. This results in a total of 23461 drifters being used, where the spatial distribution of observations is shown in \Cref{fig:spatial_dist}. Only using drogued drifters is not a restriction, it would be straightforward to simply use the data from non-drogued drifters if a practitioner was interested in a species or object which experiences a high wind forcing, or a combination of both if it is believed that the species followed a mixture of near surface and wind-forced currents. The data is quality controlled and interpolated to six hourly intervals using the methodology from \citet{Hansen1996}. These interpolated values do contain some noise due to both satellite error and interpolation, however, the magnitude of this noise is negligible in comparison to the size of grid we use in \Cref{sec:method}. Hence, we ignore this noise and treat the interpolated values as observations. For the same reason we note that the interpolation method used is not important here, instead of the six hourly product we could use the hourly product produced by methodology proposed by \citet{Elipot2016}, or drifter data smoothed by splines as proposed by \citet{Early2020}.
	
	The value of using the Global Drifter Program is we obtain a true model-free representation of the ocean. All phenomena which act on the drifters are accounted for in the data set. The other common approach is to first obtain an estimate of the underlying velocity field, then simulate thousands of trajectories using the velocity field. While this simulation approach is often satisfactory in some applications, the models generally do not agree completely with the actual observations.

	\subsection{Notation}
	Here we use $x^\circ, y^\circ$ to be a geographic coordinate corresponding to latitude and longitude respectively. We refer to the longitude-latitude grid system using the notation $x^\circ\times y^\circ$, which means each grid box goes $x^\circ$ along the longitude axis and $y^\circ$ along the latitude axis. We use bold font for any data which is in longitude-latitude pairs; i.e  $\textbf{r}=r_{lon}, r_{lat}$, and non-bold text for either a grid index or a single number. We use $\mathcal{S}$ to denote the set of all possible grid indices. \new{A full table of notation is given in Appendix \ref{sec:apx_table}.}
	
	\subsection{Capturing Drifter Motion}
	We define the drifter's probability density function as \[P(\mathbf{r}_1,t \given \mathbf{r_0}, t_0)\] where the drifter started at $\mathbf{r_0}\in \mathbb{R}^2$ at time $t_0$ and moved to position $\mathbf{r}_1\in \mathbb{R}^2$ at time $t$, where $\mathbf{r}_0$ and $\mathbf{r}_1$ are longitude-latitude pairs. In the absence of a model, this probability density cannot be estimated continuously from data alone. Therefore, we follow previous works which spatially discretize the problem \citep{Maximenko2012, Sebille2011,miron2019markov, Rypina2017, lumpkin2016fulfilling}. Instead of considering $\mathbf{r}_0\in R^2$, we consider $r_0\in \mathcal{S}$ where $\mathcal{S}$ is some set of states which correspond to a polygon in space; we will define how these are formed in Section \ref{ssec:SpatialIndex}. Often these states are simply $1^\circ\times 1^\circ$ degree boxes (e.g. as used in \Cref{fig:spatial_dist}). As in \citet{Maximenko2012}, we assume that the process driving the drifter's movement is temporally stationary. That is:
	\begin{equation}
	P(r_1,t \given r_0,t_0) = P(r_1\given r_0, t-t_0), \quad r_0, r_1\in \mathcal{S}, \nonumber
	\end{equation}
	i.e. the probability of going from ${r}_0$ to ${r}_1$ depends only on the time increment. The probability does not depend on the start or finish time. 
	
	Furthermore, given that we are using data which is observed at regular and discrete times, we shall only consider discrete values of time. Let $\mathbf{s}=\{s_0,s_1,s_2,\dots, s_n\}$ be a sequence of locations equally spaced in time where each entry $s_i$ can take the value of anything within $\mathcal{S}$. We define the probability $p(s_{i+1}=q \given s_i=k)$ as the probability that the position at time $i+1$ is $q$ given that the state at time $i$ was $k$ where $q,k  \in \mathcal{S}$. 
	
	A Lagrangian decorrelation time causes the drifter to `forget' its history~\citep{LaCasce2008}. We aim to choose a quantity which is globally higher than the Lagrangian decorrelation time. The reasoning behind using this time is that if we consider a sequence of observations, which are at least the Lagrangian decorrelation time apart then the following Markov property is satisfied: 
	\begin{align}
	&p(s_{i+1} = q_{i+1}| s_{i}=q_{i}, s_{i-1}=q_{i-1},\dotsm ,s_0= q_0) \nonumber\\ =&p(s_{i+1}=q_{i+1}\given s_{i}=q_i), \label{eq:markov}
	\end{align}
	where $q_i$ is just some fixed state and $s_{i}$ is the random process. In other words, the Markov property states that probability of transition to state $s_{i+1}$ is independent of all the past states at times $i-1$ and earlier, given the state at time $i$ is known. In this case, the physical time difference associated with $i+1$ and $i$ being larger than the chosen Lagrangian decorrelation time validates the use of the Markov assumption.

	For the rest of this paper we assume that the time between discrete time observations is equal to $\mathcal{T}_L$. We call this quantity the Lagrangian cut off time. Setting $\mathcal{T}_L$ higher than the decorrelation time allows us to use the Markov property from \Cref{eq:markov} freely. In so doing, alongside the simplification of discretizing locations, this allows the problem to be treated as a discrete time Markov chain. Here we fix $\mathcal{T}_L=5$ days as this matches previous similar works \citep{Maximenko2012, miron2019markov}. The estimated decorrelation time for the majority of the surface of the Ocean is likely to be lower than 5 days (e.g. see \citet{zhurbas2004drifter} for the Pacific and \citet{Lumpkin2002} for regions in the Atlantic). In Appendix \ref{ssec:apx_sens} we conduct a sensitivity analysis to show our results are not overly sensitive to the choice of $\mathcal{T}_L$ as long as $\mathcal{T}_L>2$ days.


	\section{Method for Computing the Most Likely Path and Travel Time}\label{sec:method}
	\citet{Maximenko2012} and \citet{VanSebille2012} focus on the use of a transition matrix estimated from drifters to discover points where drifters are likely to end up. In this section we build on such an approach by providing a method to take such a matrix and provide an ocean pathway and travel time.
	
	In Section \ref{ssec:transmatrix}, we explain in detail how the transition matrix is formed. As a grid system is needed to form the discretization of data we introduce our chosen system in Section \ref{ssec:SpatialIndex}. Then in Section \ref{ssec:likely}, we describe how we estimate the most likely path of a drifter to have taken. Finally, in Section \ref{ssec:ttest} we explain how we turn the most likely path and transition matrix into an estimate of travel time. We give a summary of how this articulates in the pseudo-code in Algorithm \ref{alg:method}.
	\subsection{Transition Matrix}\label{ssec:transmatrix}
	The location of a drifter at any given time is a continuous vector in $\mathbb{R}^2$, the longitude and latitude of the point. We define an injective map which maps this continuous process onto a discrete set of states which are indexed by integers in $\mathcal{S}$. We define the map as follows:
	\begin{equation}
	f : \mathbb{R}^2 \rightarrow \mathcal{S}.\label{eq:gridfn}
	\end{equation}
	We aim to make a Markov transition matrix $T$ of size $n_{states}$ rows and columns, where $T_{s,q}$ denotes, the probability of moving from $s$ to $q$ in one time step. Similarly to the approach of \citet{Maximenko2012}, we form our transition matrix using a gap method. In each drifter trajectory we only consider observations as a pair of points $\mathcal{T}_L$ days apart. When using this method for other applications we advise using $\mathcal{T}_L$ to be higher than the decorrelation time of velocity to justify the Markov assumption.
	
	Consider a trajectory as a sequence of positions $\mathbf{y}_j=\{\mathbf{y}_{i,j}\}_{i=1}^{n_{j}}$ where $j$ is the $j^{th}$ out of $N$ trajectories, $n_{j}$ is the number of location observations in the trajectory, and $\mathbf{y}_{i,j}\in \mathbb{R}^2$ are the longitude-latitude positions. 
	First, we map each trajectory into observed discrete states. We will denote these states as follows,
	\begin{equation}
	g_{i,j} = f(\mathbf{y}_{i,j}).\label{eq:discrete}
	\end{equation}
	For each $s, p\in\mathcal{S}$ we estimate the relevant entry of our transition matrix $T$ through using the following empirical estimate:
	
	\begin{equation}
	T_{s,p} =\dfrac{\sum_{j=1}^N\sum_{i=1}^{n_{j}-4\mathcal{T}_L} \mathbb{I}[g_{i+4\mathcal{T}_L,j}=p]\mathbb{I}(g_{i,j}=s)}{ \sum_{j=1}^N\sum_{i=1}^{n_{j}-4\mathcal{T}_L} \mathbb{I}[g_{i,j}=s]},\label{eq:tmatrix}
	\end{equation}
	\new{Where $\mathbb{I}$ is the indicator function, such that it takes the value 1 if the statement inside it is true, and zero otherwise.} Note that we take gaps of $4\mathcal{T}_L$ as observations are every 6 hours in the GDP application and $\mathcal{T}_L$ is in days. The estimation of the transition matrix, using the discretization of trajectories in \Cref{eq:discrete}, in combination with \Cref{eq:tmatrix}, is commonly referred to as Ulam's approach \citep{UlamStanislawM1960Acom}.
	We expect that states in $\mathcal{S}$ which are not spatially close will have non-zero entries such that the matrix $T$ will be very sparse, but this is not a problem for the methodology to work over large distances as we shall see.

	\subsection{Spatial Indexing}\label{ssec:SpatialIndex}
	Clearly the resulting transition matrix described in Section \ref{ssec:transmatrix} strongly depends on the choice of grid function in \Cref{eq:gridfn}. Most previous works \citep{VanSebille2012,Maximenko2012,Rypina2017, mcadam2018surface,miron2019markov} use longitude-latitude based square grids where all grid boxes typically vary between $0.5^\circ\times0.5^\circ$ and $1^\circ \times 1^\circ$. A $1^\circ \times 1^\circ$ grid cell around the equatorial region will be approximately equal area to a $111.2\text{km} \times111.2\text{km}$ square box. However, if we take such a grid above $60^\circ$ latitude, e.g. the Norwegian sea, the grid cell will be approximately equal area to a $55.6\text{km} \times 111.2\text{km}$ square box.
	
	There are a few other choices which we argue are more suitable for tracking moving data on the surface of the Earth. Typically three types of grids exist for tessellating the globe: triangles, squares, or a mixture of hexagons and pentagons. Here we choose to use hexagons and pentagons as they have the desirable property that every neighbouring shape shares precisely two vertices and an edge. This is different to say a square grid where only side-by-side neighbours share two vertices and an edge, whereas diagonal neighbours share only a vertex. This equivalence of neighbors property for hexagons and pentagons is clearly desirable for the tracking of objects as this will result in a smoother transition matrix.
	
	\begin{figure}[ht]
		\centering
		\includegraphics[width=228pt]{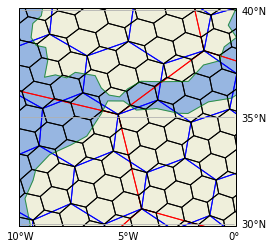}
		\caption{A small area around the Strait of Gibraltar which is tessellated using the H3 spatial index. We show resolutions 1, 2 and 3 in red, blue and black respectively. Black is the resolution used in this work.}
		\label{fig:h3_index}
	\end{figure}
	We specifically use the grid system called \textit{H3} by UBER \citep{H3}.  This system divides the globe such that any longitude and latitude coordinate is mapped to a unique hexagon or pentagon. This shape will have a unique \textit{geohash} which we can use to keep track of grid index. The benefit of using such a spatial indexing system is that attention is paid to ensuring that each hexagon is approximately equal area. We use the {\em resolution 3} index where each hexagon has an average area of $12,392 \text{km}^2$. A square box of size $111.32\text{km}\times 111.32\text{km}$ has roughly the same area as this which is very similar to the size of a $1^\circ\times 1^\circ$ grid cell near the equator. An example of an area tessellated by H3 is shown in \Cref{fig:h3_index}. Other potential systems which could be used include S2 by Google which is a square system, or simply using a longitude-latitude system as various other works do. \new{We show some example pathways using different grid systems and resolutions in the Supplementary Information Figure S1. The longitude-latitude system results in pathways that unrealistically follow long block-wise vertical or horizontal straight line motions, in contrast to the more realistic and meandering pathways produced by the hexagonal-pentagonal \textit{H3} grid system.}
	
	\subsection{Most Likely Path}\label{ssec:likely}
	For our analysis, the first step is to define a most likely path. A path is simply a sequence of states such that the first element is the origin and the last element is the destination. We also require that two neighboring states are not equal to each other.
	\begin{mydef}[Path]
		We define the space of possible paths $\mathcal{P}_{o,d}$, between the origin $o\in \mathcal{S}$ and destination $d\in\mathcal{S}$, as the following:
		\begin{align*}\mathcal{P}_{o,d}&=\{\mathbf{p}=(p_0, p_1, p_2,\dots, p_n) : p_i \in \mathcal{S}\\
		&  \forall i \in\{1,\dots,{n-1}\},\ p_0=o,\ p_n=d, p_{i-1}\ne p_i\}.
		\end{align*} 
		With a cardinality operator $|\mathbf{p}|=n$ which denotes the length of the path.
	\end{mydef} 
	Given the transition matrix $T$ we define the probability of such a path: 
	\begin{equation}
	P(\mathbf{p}) = \prod_{i=0}^{n-1} P(s_{i+1}=p_{i+1}\given s_i=p_i) = \prod_{i=0}^{n-1} T_{p_i, p_{i+1}}.
	\end{equation}
	
	\begin{mydef}[Most likely path]
		Consider any path $\mathbf{p}\in \mathcal{P}_{o,d}=\{p_0, p_1, p_2,\dots, p_n\}$. By the most likely path $\hat{\mathbf{p}}$ we mean the path which maximises the probability of observing that path.
		\begin{equation}
		\mlpath=\argmax_{\mathbf{p}\in \mathcal{P}_{o,d}}\left\{ P(\mathbf{p})\right\}= 
		\argmax_{\mathbf{p}\in \mathcal{P}_{o,d}} \left\{\prod_{i=0}^{n-1}T_{p_i,p_{i+1}}\right\}\label{eq:most likely}.
		\end{equation}
	\end{mydef}
	Optimising \Cref{eq:most likely} appears intractable at first glance. \new{However, this can easily be solved with shortest path algorithms such as Dijkstra's algorithm~\citep{dijkstra1959note}. We give precise details on how to find this pathway in Appendix \ref{sec:apx_shortest}.}
	
	\subsection{Obtaining a travel time estimate}\label{ssec:ttest}
	The most likely path is often a quantity of interest in itself, however we can also obtain a travel time estimate of this path. The method should be fast and efficient as it should be able to run for large sets of locations quickly. We achieve this by giving a formula to estimate the travel time based directly on the transition matrix.

	Consider the path, $\mathbf{p}= \{p_1, \dots, p_n\}$, from which we aim to estimate the expected travel time. The key consideration this section addresses is: the path is a sequence of unique states, whereas when simulating from a discrete time Markov chain, the chain has a probability of remaining within the same state for multiple time steps. We therefore aim to obtain an estimate of how long the Markov chain takes, on average, to jump between $p_i$ and $p_{i+1}$, and then aggregate this over the path to form a travel time estimate.
	
	We assume that the only possibility is that the drifter follows the path we are interested in. So $p_i$ must be followed by $p_{i+1}$. Now we use $t$ to index the time of the Markov chain and suppose $s_t=p_i$. We are then interested in the random variable $k$ where $t+k$ is the first time that the process transitions from $p_i$ to $p_{i+1}$. Note that the only possibility for states $\{s_{t+l}\}_{l=1}^{k-1}$ is that they are all $p_i$, otherwise the object would not be following the path of interest. Therefore, we obtain the distribution of $k$ as follows (proof in Appendix \ref{ssec:apx_deriv}):
	\begin{align}
	&P(s_{t+k}=p_{i+1}, \{s_{t+l}=p_i\}_{l=1}^{k-1}\given s_{t}=p_{i}, \mathbf{p}\})\nonumber\\
	&=\dfrac{T_{p_{i}, p_{i+1}} T_{p_i,p_i}^{k-1}}{(T_{p_i,p_i}+T_{p_i,p_{i+1}})^{k}}.\label{eq:negbinom}
	\end{align}
	Note that if we set $a=\dfrac{T_{p_{i}, p_{i}}}{T_{p_i,p_i}+T_{p_i,p_{i+1}}}$ in \Cref{eq:negbinom} we get:
	\begin{equation}
	P(s_{t+k}=p_{i+1}\given s_{t}=p_{i}, \mathbf{p})= a^{k-1}(1-a), \label{eq:negbin}
	\end{equation} which is the probability distribution function of a negative binomial distribution with success probability $a$ and number of failures being one. We denote the random variable for the travel time between $p_{i}$ and $p_{i+1}$ as $k_i$. As the negative binomial distribution corresponds to the time until a failure, we are interested in taking one time increment longer than this as we require $k_i$ to be the time that we move from $p_i$ to $p_{i+1}$ i.e. the time of the failure. Therefore the distribution of $k_i$ exactly follows $k_i-1\sim  \text{NB}(1,a)$. Also, note that $k_i$ is in units of the chosen Lagrangian cutoff time $\mathcal{T}_L$.
	
	To get the expectation of the total Lagrangian travel time we consider the sum of all the individual parts of the travel times $\mathbf{k}=\sum_{i=0}^{n-1} k_i$, such that we obtain:
	\begin{equation}
	\Ex[\mathbf{k}]=\sum_{i=0}^{n-1}\Ex[k_i]= \sum_{i=0}^{n-1} \left(\dfrac{T_{p_{i}, p_{i}}}{T_{p_{i}, p_{i+1}}}+1\right), \label{eq:exp}
	\end{equation}
	where we have used that the expectation of the negative binomial is $\Ex[x \sim \text{NB}(1,a)] = \frac{a}{1-a}$.
	
	We could attempt to obtain a simple variance estimate for the estimate $\Ex[\mathbf{k}]$ with classical statistics. However, we would only be able to account for variability within the estimates of the entries $T$, as we would have to assume $\mathbf{p}$ is known. In our case we are interested in the time of $\mlpath$, which is itself an estimate as it depends on $T$. Obtaining any analytical uncertainty in the estimation of the most likely path would be intractable due to the complexity of the shortest path algorithm. Therefore, we propose to address the issue of uncertainty in $\Ex[\mathbf{k}]$ and $\mathbf{p}$ due to data randomness in Section \ref{ssec:boot} using the non-parametric bootstrap. To finish this section, we provide the pseudo-code for our approach in Algorithm~\ref{alg:method}.
	\begin{algorithm}
		\SetAlgoLined
		\DontPrintSemicolon
		\KwIn{Drifter data set $\mathbf{y}$, a set of locations $\mathbf{x}$, Lagrangian cutoff time $\mathcal{T}_L$}
		
		Map all the drifter locations $\mathbf{y}$ to their grids $g_{j,i}=f(\mathbf{y}_{j,i})$ using the map from \Cref{eq:gridfn}. \;
		Map all the locations of interest to their grids $g^{x_i}=f(x_i)$.\;
		
		Form transition matrix $T$ using \Cref{eq:tmatrix}.\;
		\For{each unique pair $o$ and $d$ in $\{g^{x_i}\}_{x_i \in \mathbf{x}}$}{
			Find and store the most likely path $\mlpath_{o,d}$ \new{by optimizing Equation~\eqref{eq:most likely}}.\;
			Using this path, find and store the expected travel time $\Ex[\hat{\mathbf{k}}_{o,d}]$ using \Cref{eq:exp}.\;
		}
		\KwResult{Travel times $\Ex[\hat{\mathbf{k}}_{o,d}]$ for every pair of locations in $\mathbf{x}$ and a corresponding path $\mlpath_{o,d}$ given as a sequence of grid indices in $\mathcal{S}$.}
		\vspace{0.5cm}
		\caption{Pseudo-code which summarises how \Cref{sec:method}  is used to turn drifter data and a spatial index function into most likely paths and travel time estimates.}
		\label{alg:method}
	\end{algorithm}

	\section{Stability and Uncertainty}\label{sec:stability}
	
	\subsection{Random Rotation}\label{ssec:rot}
	A key consideration is that the final results of the algorithmic approach may strongly rely on the precise grid system $f$ chosen in \Cref{eq:gridfn}. To address the uncertainty due to the discretization we propose to {\em randomly sample} a new grid system then run the algorithm for the new grid system. In a simple 2d square grid one could simply sample a new grid system by sampling two numbers between 0 and the length of a side of the square, then shifting the square by these sampled amounts in the $x$ and $y$ direction. In global complicated grid systems such as the one we consider here proposing uniform random shifting is not trivial.
	
	Rather than trying to reconfigure the grid system, instead we suggest a more universal alternative. We suggest randomly rotating the longitude-latitude locations of all the relevant data using random rotations. Such a strategy will work for any spatial grid system as it just involves a prepossessing step of transforming all longitude-latitude coordinates\footnote{Conditional on the grid system having a reasonable minimum area. This method rotates the poles to a random point, which would give spurious results in a longitude-latitude grid. Thus providing another reason why the H3 system is more suitable.}. Note that for each rotation we are required to re-assign the points to the grid and re-estimate the transition matrix. These are the two most computationally expensive procedures of the method. To generate the random rotations we use the method suggested by \citet{shoemake1992uniform}. In summary, it amounts to generating 4 random numbers on a unit 4 dimensional hypersphere as the quaternion representation of the 3 dimensional rotation, which can equivalently be represented as a rotation matrix $M$. Then we apply this rotation to the Cartesian representation of longitude and latitude. 

	To obtain travel times which remove bias effects from discretization, we sample $n_{rot}$ rotation matrices $M^{(i)}$. We then run Algorithm \ref{alg:method}, however as a prepossessing step we rotate all locations of the drifter trajectories and locations of interest. For each rotation matrix this will result in a set of travel times $\hat{d}^{(i)}$. The sample mean of these rotations will be more stable than the vanilla method. The sample standard deviation will inform us about uncertainty in travel times due to discretization.
	\subsection{Bootstrap}\label{ssec:boot}
	If we required a rough estimate of uncertainty we could consider that $\mlpath$, the most likely path, is fixed and then estimate $\Var[\hat{\mathbf{k}}]$. However, this would be a poor estimate because such an estimate would assume that: (1) the transition matrix entries follow a certain distribution, and (2) the path $\mlpath$ is the true most likely path. In reality neither of these are true, they will both just be estimates. The transition matrix elements are estimated from limited data and the shortest path strongly depends on the estimated transition matrix, e.g. a small change in the transition matrix could result in a significantly different path. Therefore, we obtain estimates of uncertainty by bootstrapping \citep{EfronBradley1993Aitt}.
	
	Bootstrapping is a method to automate various inferential calculations by resampling. Here the main goal is to estimate uncertainty around $\hat{\theta} = \Ex[\hat{\mathbf{k}}]$. The bootstrap involves first resampling from the original drifters to obtain a new data set. We call  $\mathbf{y}^*=\{\mathbf{y}_{j}^*\}_{j = 1, \dots N}$ a bootstrap sample, where $\mathbf{y}_j^*$ is a drifter trajectory which has been sampled with replacement from the original $N$ drifters. Then we use $\mathbf{y}^*$ as the input dataset to Algorithm~\ref{alg:method}.
	
	We do this resampling $B$ times to obtain $B$ estimates of $\hat{\theta}=\Ex[\mathbf{\hat{k}}]$, we denote these bootstrap estimates as $\{\hat{\theta}^{(b)}\}_{b=1}^B$. We then estimate our final bootstrapped mean and standard deviation estimates as the following:
	\begin{align}
	sd_{boot}^2&=\left[\dfrac{\sum_{b=1}^B \left(\hat{\theta}^{(b)}-\hat{\theta}^{(.)}\right)^2}{B-1}\right], \nonumber\\
	\text{where}\quad \hat{\theta}^{(.)}&=\sum_{b=1}^B\hat{\theta}^{(b)}/B.\label{eq:bootstrapse}
	\end{align}
	
	In addition to the uncertainty measure in travel time that both the bootstrap and rotation methodology provide, these methods also supply a collection of sample most likely paths. These paths can be used to investigate various phenomena, such as why the uncertainty is high. We can plot the paths for a fixed origin-destination pair and may see for example that many paths follow one current where another collection of paths follow a different current. We give numerous examples of this in Sections \ref{ssec:res_boot} and \ref{ssec:res_rot}.

	\section{Application}\label{sec:results}
	\begin{table}[bth]
		\centering
		\begin{tabular}{lrr}
			\toprule
			{} &  Longitude &  Latitude \\
			\midrule
			1 &        9.0 &     -25.5 \\
			2 &      -25.0 &      -5.0 \\
			3 &      -45.0 &     -40.0 \\
			4 &      -69.0 &      39.0 \\
			5 &      -42.5 &      41.5 \\
			6 &      -42.0 &      27.5 \\
			7 &      -93.2 &      24.8 \\
			\bottomrule
		\end{tabular}
		\caption{Table of station locations}
		\label{tbl:locations}
	\end{table}
	
	We use the locations given in \Cref{tbl:locations} for the demonstration of the method described in this paper. These locations were chosen for multiple reasons; (1) they were placed on or near ocean currents, such as the South Atlantic current, the Equatorial current and the Gulf Stream; the magnitudes of which can be seen in \Cref{fig:Station_loc}, and (2) stations were placed in both the North and South Atlantic to show how the method can find pathways which are not trivially connected. First we go over an application of the vanilla method from \Cref{sec:method}, then we provide brief results using the adaptations using bootstrap and rotations from \Cref{sec:stability} in Section \ref{ssec:res_boot} and Section \ref{ssec:res_rot} respectively. We supply a link to a Python package and code used to create these results in the Appendix.
	
		\begin{figure*}[ht!]
		\centering
		\includegraphics[width=\textwidth]{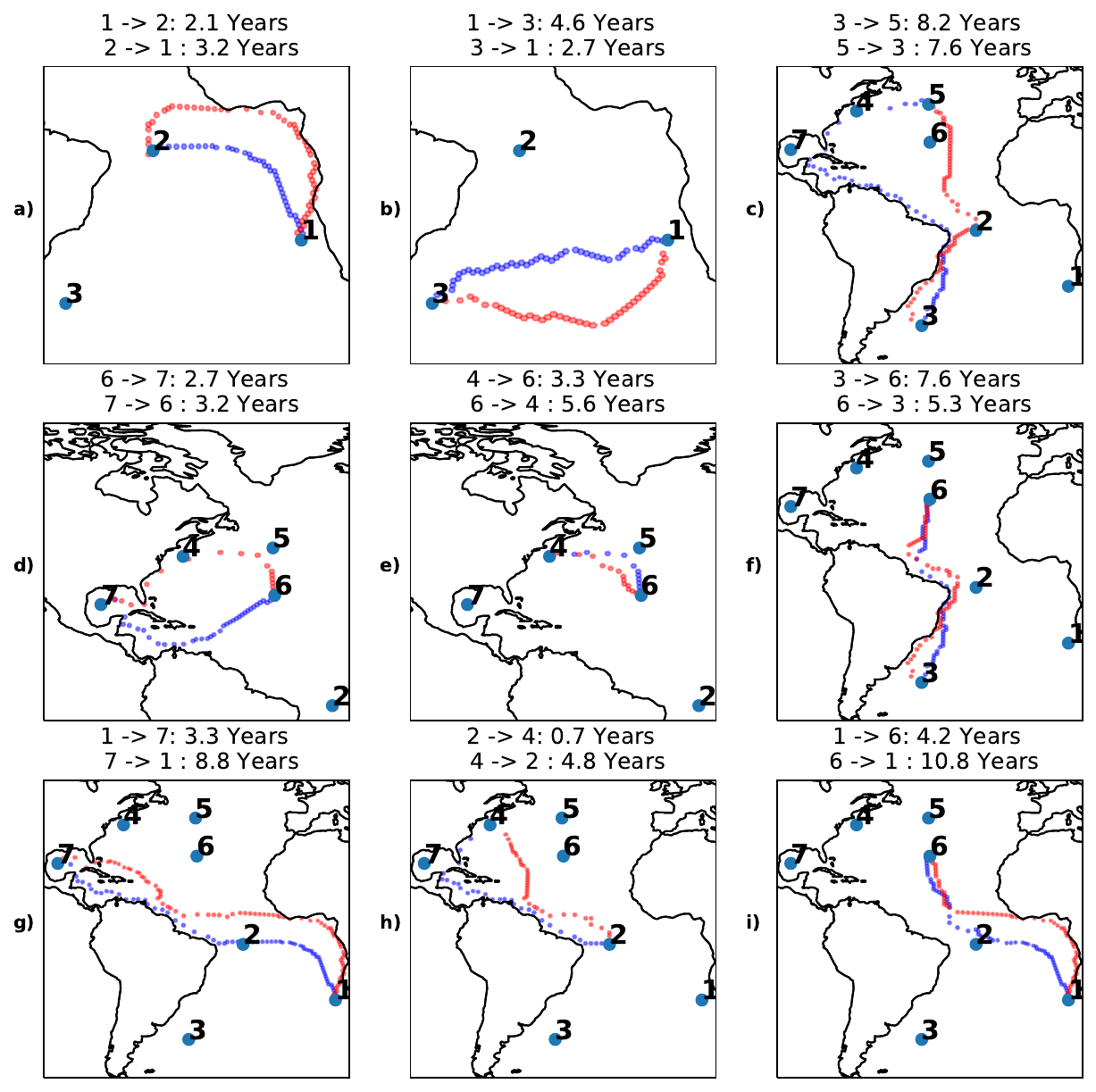}
		\caption{Example pathways found from the method. Sequences of blue hexagons are going from the lower number to the higher number. Sequences of red hexagons are going from the higher number to the lower number. Numbered locations are as in \Cref{tbl:locations}. The expected travel time of the most likely path is given in the title of each plot. Similar plots can be provided for every location pair using the online code, however these are not presented here owing to page length considerations.}
		\label{fig:likely_pathway}
	\end{figure*}
	
	Prior to our analysis we take a practical step to improve the reliability of the method. we find the states corresponding to $-79.7^\circ, 9.07^\circ$, $-80.73^\circ,8.66^\circ$ (two points on the Panama land mass), $-5.6^\circ, 36^\circ$ and $-5.61^{\circ}, 35.88^\circ$ (two points on the Strait of Gibraltar), then remove the corresponding rows and columns from $T$. If this step is not taken the method often uses pathways crossing the Panama land mass, resulting in impossibly short connections to the Pacific Ocean. The reasoning for removing the points on the Strait of Gibraltar is data-driven, further details are in the supplementary information, particularly how one can adapt the method to specify travel times into and out of the Mediterranean sea.
	

	\Cref{fig:likely_pathway} shows the pathways between a representative sample of the stations.
	First we note what features are observed in the most likely path. The Gulf Stream is used on almost every path trying to access locations 4, 5 or 6 in \Cref{fig:likely_pathway}. Observe in \Cref{fig:likely_pathway} $c)$ when going from location 3 to 5 that the method chooses to enter the Gulf of Mexico and then uses the Gulf Stream to access location 5, even though the actual geodesic distance of this path is long. Other examples which use the Gulf Stream include $d)$ and $h)$. Generally, any of the paths leaving location 1 and attempting to travel northwest use the Benguela Current, for example \Cref{fig:likely_pathway} $a)$, $i)$ and $g)$.
	
	The travel times obtained between the sample stations in \Cref{fig:likely_pathway} show interesting results regarding the lack of symmetry when reversing the direction of the path between two stations. When going from location 2 to location 4 we estimate a long most likely path in terms of physical distance. However, the resulting travel time of this path (0.7 years) is smaller than the travel time of the more direct path from location 4 to location 2 (4.8 years) - which is much shorter in distance. This is because the path going from location 2 to location 4 follows strong currents such as the North Equatorial current and the Gulf Stream. Another interesting result is that going from 3 to 5 and vice versa are relatively close in terms of travel time even though 3 to 5 uses the Gulf Stream but the return does not. In the most likely path from 3 to 5, up until around $-16^\circ$ latitude the travel time is 5.2 years, which we expect as the pathway seems to be going against the Brazil current. After this point the rest of the path takes the remaining 3 years despite the remainder being over half the actual physical distance of the pathway. We expect this short time is due to the method finding a pathway along the North Brazil current, followed by the Caribbean current, followed by the Gulf Stream.
	
	\begin{figure}[ht!]
		\centering
		\includegraphics[width=228pt]{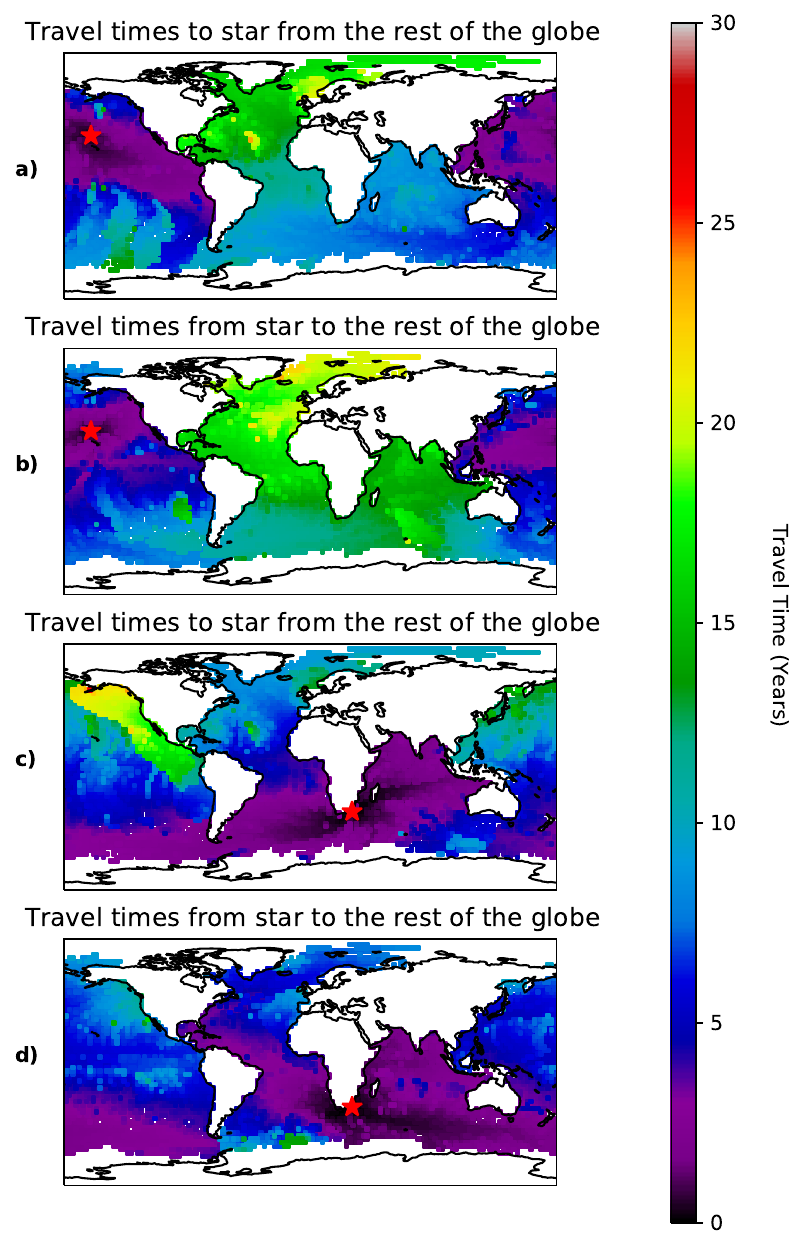}
		\caption{Travel times of the most likely path originating from the red stars and going to or from (indicated by the title) the centroid of a $2.5^\circ\times 2.5^\circ$ square grid system. Figure setup and locations taken to match Figure 2 of \cite{jonsson2016timescales}.}
		\label{fig:one_to_all}
	\end{figure}
	\subsection{Global Travel Times}
	\label{ssec:global_travel_times}
	\Cref{fig:one_to_all} shows the travel time distribution to and from two fixed locations, taken to match the studied locations of \cite{jonsson2016timescales}, to the entire globe. We note that the travel time map is less smooth than the one shown in \cite{jonsson2016timescales}. The black and purple areas however (up to 5 years travel time) are similar to those found in \cite{jonsson2016timescales}, showing agreement over short spatial scales. When it comes to larger distances we generally find the maps are markedly different. For example the yellow patch in the north east pacific in \Cref{fig:one_to_all}c is not seen in \cite{jonsson2016timescales}. Such discrepancies can be attributed to many reasons such as: (1) they reflect the difference in methods, where we use a transition matrix approach, and \cite{jonsson2016timescales} use a connectivity matrix; (2) \cite{jonsson2016timescales} aim to find the shortest path in time, whereas we aim to find the expected time of the most likely path; and (3) the results shown here are derived from real data, whereas \cite{jonsson2016timescales} use simulated trajectories. 
	
\begin{figure}[ht]
		\centering
		\includegraphics[width=228pt]{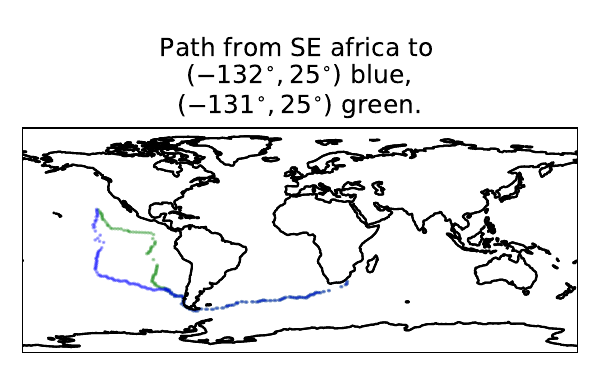}
		\caption{The most likely path from two points in the North Pacific to the south-east coast of Africa. The green and blue pathways are almost identical as they cross the south Atlantic. The pathways differ greatly however as they cross the Pacific, even though the two starting points in the north Pacific are only 1 degree apart. The path going from $-131^\circ,25^\circ$ has an expected travel time of 21.2 years, the path going from $-132^\circ,25^\circ$ has an expected travel time of 11.4 years.}
		\label{fig:different paths}
	\end{figure}

	We show an example in Figure \ref{fig:different paths} which explains the lack of spatial smoothness in \Cref{fig:one_to_all}, where we show two pathways both originating from a fixed point and ending at two distinct points only $1^\circ$ latitude apart. The points are on either side of the discontinuity in the north-east Pacific seen in Figure \ref{fig:one_to_all}c. The pathways become visibly different after they have both reached the south Pacific. Such a phenomenon results in the lack of spatial smoothness of travel time distributions. This demonstrates that the travel times do not necessarily obey the triangle inequality. If smoothness is desired we show an alternative approach in the supplementary information, where instead a minimum travel time is the objective, which is then more analogous to the \cite{jonsson2016timescales} approach. We argue however that the expected travel time of the most likely path, rather than the minimum travel time, is a more relevant metric for estimating connectivity and Lagrangian distance in applications measuring spatial dependence between points in the ocean.

	\subsection{Bootstrap}\label{ssec:res_boot}
\begin{figure}[th]
		\centering
		\includegraphics[width=228pt]{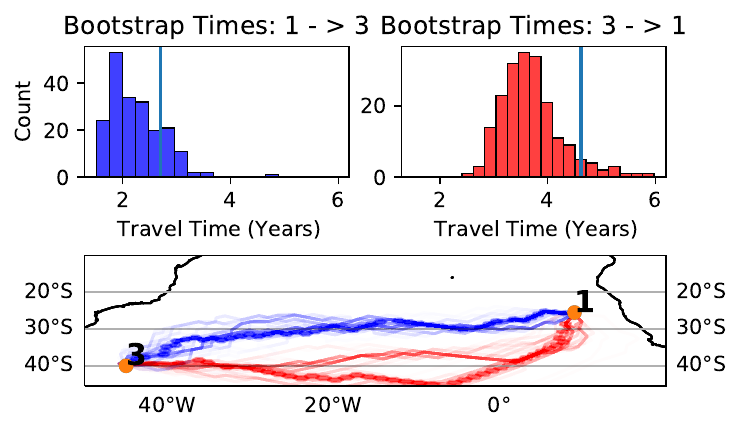}
		\caption{Two bootstrap distributions of travel times are shown in the top row resulting from 200 bootstrap samples. The vertical line is the travel time if the full data is used to estimate the path and time. The corresponding bootstrapped paths are shown in the bottom figure. Blue lines and hexagons are for going from $1$ to $3$, red lines and hexagons are for going from $3$ to $1$. The lines connect the centroids of the spatial index of the bootstrapped paths. Darker lines mean that path is taken more often. The light hexagons are the pathway taken if the full data is used with no resampling i.e.\ the pathway shown in \Cref{fig:likely_pathway}.}
		\label{fig:bootstrap}
	\end{figure}

	To show the value of the bootstrap we show the results for one particular pair of stations, the pathway going from location 1 to location 3 and back. The pathways which result from the bootstrap are shown in the bottom panel of \Cref{fig:bootstrap}. The darker lines on the figure imply that that this transition is used more often. We see that for most of the journey the darker lines closely follow the original path. The bootstrap discovers some slightly different paths, for example around $-20^\circ$ Longitude the path going from 3 to 1 occasionally seems to find that going further south is a more likely path. Also, around the beginning of the path going from 1 to 3, we see that the most likely path taken most frequently by the bootstrap samples often does not follow the most likely path from the full data.
	
	The main goal of the bootstrap is that we obtain an estimate of the standard errors. In this case we get standard error estimates using \Cref{eq:bootstrapse} of 0.5 years for going from 3 to 1 and 0.6 years for going from 1 to 3. In general, we found that the standard error was lower when the path follows the direction of flow. The top row of plots in \Cref{fig:bootstrap} appears to show that there is a slight bias between the bootstrap mean and the vanilla method travel time. We believe this is due to the variance within the paths. The mean estimated from the bootstrap samples are close to the estimates from the rotation method we will shortly present in \Cref{fig:pairwise_rot}. The rotation mean estimates are within 0.4 years of the bootstrap means in both cases shown here.
	
	\subsection{Rotation}
	\label{ssec:res_rot}

\begin{figure}[t]
		\centering
		\includegraphics[width=228pt]{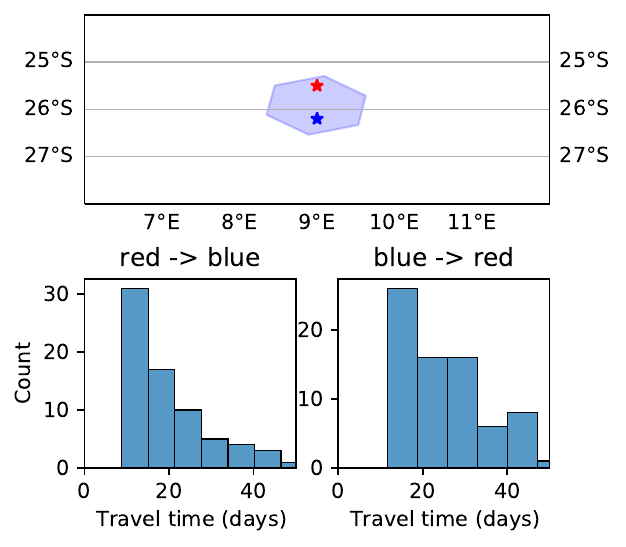}
		\caption{Plot of location 1 from \Cref{tbl:locations} and the point $9^\circ, -26.2^\circ$, which is $0.7^{\circ}$ south of location~1. The relevant H3 hexagon is plotted over the points. In the bottom row we plot the histogram and density estimate of the travel times in each direction from applying 100 rotations. The 22 zeros for when the two locations are in the same hexagon are not included in the histogram.}
		\label{fig:Samegridindex}
	\end{figure}

	If we consider two points in the same H3 Index, for example location 1 ($9^\circ,-25.5^\circ$) and a new point $9^\circ, -26.2^\circ$ (as shown in \Cref{fig:Samegridindex}), then using the original grid system the method will simply produce a travel time of 0. To solve this problem, we consider using 100 rotations as explained in Section \ref{ssec:rot}. For each rotation we estimate the travel time both back and forth. In 22 of the rotations the two points ended up in the same hexagon, hence resulting in a zero travel time. We plot the distribution of the other 78 travel times in the bottom row of \Cref{fig:Samegridindex}. The mean of all the entries including the zeros is 20.5 days for going from the new point to location 1, and 22.2 days for going from location 1 to the new point. 
	\begin{figure*}[!ht]
		\centering
		\includegraphics[width=\textwidth]{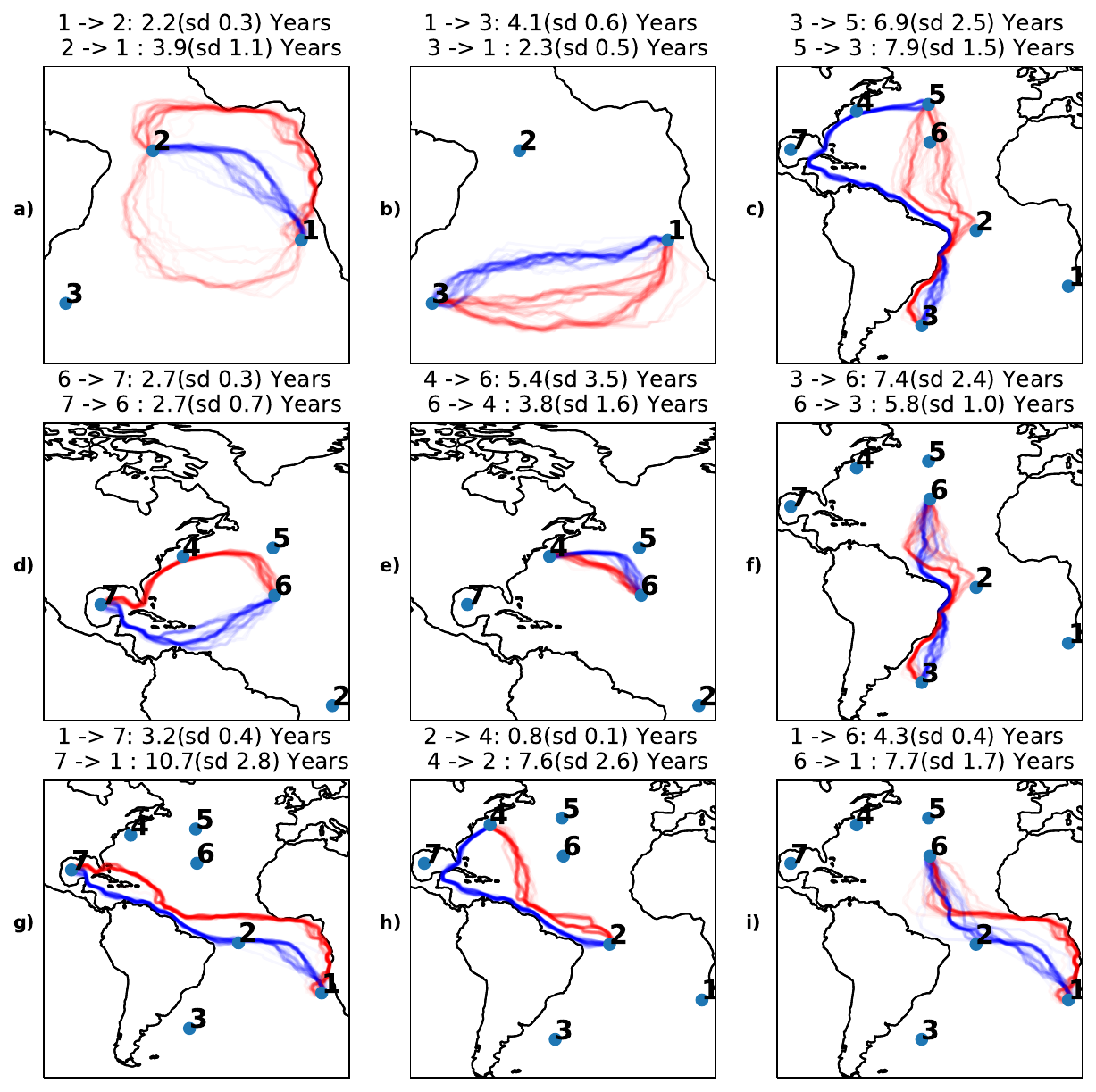}
		\caption{This figure layout is the same as in \Cref{fig:likely_pathway}, except here we plot paths resulting from 100 random rotations. Each line connects the centroid of each hexagon within the path. Note that the hexagons now come from rotated grid systems, so the centroids could be at any location hence the smooth continuous looking lines. The lines are plotted with transparency, when multiple lines overlap these lines will look darker. Standard deviations of the travel times of the 100 paths are reported in the title of each figure.}
		\label{fig:pairwise_rot}
	\end{figure*}
	
	The second benefit of performing rotations is to make estimates less dependent on the grid system. We use the same 100 rotations as with the previous example, and compute the most likely path and the mean travel times. In \Cref{fig:pairwise_rot} we plot the pathways with the mean and standard deviation of the travel times resulting from these 100 rotations. The travel times and paths shown in this figure are comparable to those given in \Cref{fig:likely_pathway}. In most of the pathways we see that the darkest, most popular paths match up with the pathways in \Cref{fig:likely_pathway}. 
	
	One of the more interesting results from this analysis is the path going from $2$ to $1$ in \Cref{fig:pairwise_rot} a). Most of the paths go up closer to the Equator, then use the Equatorial Counter current, followed by the Guinea and Gulf of Guinea currents as in the original vanilla application of the methodology. A small number of the rotations result in pathways that end up crossing the South Atlantic, to the south of location 2, then follows the South Atlantic current over to location 1.
	
	In general, the travel times from the rotation and original method can be significantly different, which supports the need for this rotation methodology. If we compare \Cref{fig:likely_pathway} and \Cref{fig:pairwise_rot}, most of the distances stay close to what they were in the original results using no rotations. We see that going from 6 to 4 drops from 5.6 years  in \Cref{fig:likely_pathway}e) to 3.8 years in \Cref{fig:pairwise_rot}e) and 4 to 6 increases from 3.3 years to 5.4 years. This causes the ordering of the distances to change as 6 to 4 is now the shorter travel time. We believe the case in e) is mainly due to 4 being located just north west of the stronger currents of the Gulf Stream, which makes it sensitive to the grid system. However, the high standard errors in \Cref{fig:pairwise_rot} suggest we are uncertain about this travel time.

	\section{Discussion and Conclusion}\label{sec:Conclusion}
	In contrast to \citet{VanSebille2014}, our methodology as presented does not take into account seasonality. We have a few ideas for how seasonality could be incorporated in future work. An obvious adaptation, if the aim was to obtain a short travel time which is expected to lie in a small 3 month window, is to just estimate $T$ using drifter observations which are in that time window. Alternatively, we could use $\mathcal{T}_L$ to be a certain jump such as a gap of two months, then we estimate 6 transition matrices say $T^{(k)}$, where the entries $T^{(k)}_{i,j}$ are probabilities of going from the previous time period at state $i$ to state $j$ at the current time. Such a set up could still be solved using our shortest path algorithm.
	We justify our approach in the same way as \citet{Maximenko2012}: we aim to provide a global view and a simple general concept explaining the pattern of potential pathways and travel times. The base method can then be adapted by practitioners to account for local spatial or temporal considerations.
	
	More results demonstrating the robustness of our method, and motivation of parameter choices, can be found in the supplementary information. A key finding we discuss here, is that we found the size of the grid system affects the estimated travel times significantly, regardless of whether the lat-lon or the \textit{H3} grid system is used. Therefore we do not recommend comparing travel times obtained from two different grid sizes. Generally the results are correlated in an order comparison sense, however, their magnitudes change. Typically a smaller grid system results in shorter travel times. Due to this we would only advise the results to be used in relative comparison to each other, for example by saying that the travel time from $a$ to $b$ is twice that than from $b$ to $c$, where both times are obtained with the same grid system. The choice to show resolution 3 in this paper was found to perform robustly (balancing the error from discretisation and data sparsity), and follows grid sizes that approximately match previous works where $1^\circ \times 1^\circ$ grids are used, but this can be changed easily in the online package.
	
	The use of the bootstrap and rotations are relatively easy methods to implement, each of which provides effective estimates of uncertainty from data uncertainty and discretisation respectively. However, combining these procedures into one requires careful consideration. If we wanted to run $n_{rot}$ rotations and $B$ bootstraps for each rotation, we still require a method to combine these estimates of travel times. We could treat every rotation equivalently, so say that our bootstrap sample in \Cref{eq:bootstrapse} is all $n_{rot}\times B$ samples to obtain an estimate of uncertainty in travel time due to the combination of grid discretization and data randomness. Additionally, we could decompose the uncertainty and provide a standard error for just the data randomness by estimating a standard error for each rotation using just the $B$ samples in each rotation, and then taking the average of all $n_{rot}$ standard error estimates.
	
	Our choice of the Lagrangian decorrelation time of 5 days may be too low in some instances. Previous works have found correlations in the velocity data lasting longer than $5$ days in certain regions \citep{Lumpkin2002,zhurbas2004drifter,Elipot2010}. This may suggest that using a larger value for $\mathcal{T}_L$ may be needed to justify the Markov assumption. The tradeoff however is resolution, where shorter timescales allow pathways and distances to be computed with more detail. Our methodology is designed flexibly such that the practitioner can pick the most appropriate timescale for the spatial region and application of interest.
	
	In general some unexpected features of the method do occur such as the discontinuity discussed in Section \ref{ssec:global_travel_times}. We expect there would be less of a discontinuity if these times were computed with the rotation methodology, however we argue that the discontinuities with travel times of most likely pathways should always exist. If smoothness of travel times was a major requirement, then one could consider the {\em shortest} path in travel time rather than the {\em most likely} path. We briefly show this adaptation in the supplementary information. We expect the results would require more careful checking in such an approach, as the shortest path would be more likely to use any glitches in the grid system such as if there was a connection over Panama. 
	
	To summarize, in this paper we have created a novel method to estimate Lagrangian pathways and travel times between oceanic locations, thus offering a new, fast and intuitive tool to improve our knowledge of the dynamics of marine organisms and oceanic transport and global circulation.

\subsection*{Acknowledgments}
The work of M. O'Malley was funded by the Engineering and Physical Sciences Research Council (Grant EP/L015692/1). The work of A. M. Sykulski was funded by the Engineering and Physical Sciences Research Council (Grant EP/R01860X/1).
\subsection*{Data Availability Statement}
    The drifter data were provided by the Global Drifter Program \citep{GDP}. The currents used for visualisation purposes in \Cref{fig:Station_loc} are V3.05 of the dataset supplied on the Global Drifter Program website \citep{laurindo2017improved}. 

    \appendix
    \section{Package}
    Code to reproduce all figures related to the method is available at \url{https://github.com/MikeOMa/MLTravelTimesFigures} which depends on the  python package implementing all of the above methods in this paper at \url{https://github.com/MikeOMa/DriftMLP}. The package takes roughly 3 minutes total to go from raw data to a pairwise travel time matrix for the locations shown in Table \ref{tbl:locations} using Algorithm \ref{alg:method}.\\
\section{Table of Notation}
\label{sec:apx_table}

\new{We include a table of mathematical notation for reader reference in \Cref{tbl:notation}.}\\
\begin{table}

\begin{tabular}{|p{0.1\textwidth}|p{0.3\textwidth}|}
\hline
     $P(x\given y)$ & denotes the probabilities of event(s) $x$ given that $y$ occurs.\\
     \hline
     $\Ex[x]$& The expectation of $x$.\\
     \hline
     $f(s)$ & The discretization function i.e. H3. \\
     \hline
     $\mathbb{I}(x)$ & Indicator function giving 1 if $x$ is true and 0 otherwise.\\
     \hline
     $\argmax_{x \in S}$ & An operator which gives the input value, which maximises the function $q$, restricted to the set $S$.\\
     \hline
     $T$, $T_{i,j}$ & $T$ denotes transition matrix, with entries $T_{i,j}$. $i,~j \in \mathcal{S}$, denoting the probability of moving from state $i$ to $j$ in $\mathcal{T}_L$ days.\\
     \hline
     $x^\circ \times y^\circ$ & refers to a longitude-latitude grid system, $x$ degrees in the longitudinal direction, $y$ degrees in the latitudinal direction.\\
     \hline
     $\mathcal{T}_L$ & Lagrangian cut off time.\\
     \hline
     $\mathcal{S}$ & The set of all possible spatial indices.\\
     \hline
     $\mathcal{P}_{o,d}$ & The set of all possible paths going from $o$ to $d$.\\
     \hline
     $\mathbf{p}= \{p_i\}_{i=1}^n$ & A pathway of length $n$. Indicates a sequence $p_1,~ p_2, \dots, p_n$. All $p_i\in\mathcal{S}$.\\
     \hline
     $\mathbf{k}$ & The expected travel time of a path $\mathbf{p}$. \\
     \hline
      $\hat{\mathbf{p}}$, $\hat{\mathbf{k}}$ & Hat notation implies we are considering the most likely path and travel time of that path respectively.\\
      \hline
      $s_t$ & used to index the state of the Markov chain after $t$ steps.\\
      \hline
\end{tabular}
\caption{Table of mathematical notation.}
\label{tbl:notation}
\end{table}
\section{Finding the shortest path}
\label{sec:apx_shortest}
\new{To solve the optimization of \Cref{eq:most likely}, we can equivalently consider the $log$ of $P(\mathbf{p})$:
	\begin{align}
	\log P(\mathbf{p}) &= \sum_{i=0}^{n-1}\log T_{p_i,p_{i+1}}.\nonumber
	\end{align}
	Then we use the fact that:
	\begin{align}
	\hat{\mathbf{p}}& = \argmax_{\mathbf{p}\in \mathcal{P}_{o,d}}\left\{ \log P(\mathbf{p})\right\} = \argmin_{\mathbf{p}\in \mathcal{P}_{o,d}} \left\{- \log P(\mathbf{p}\right\}) \nonumber\\
	&= \argmin_{\mathbf{p}\in \mathcal{P}_{o,d}}\left\{ -\sum_{i=0}^{n-1}\log T_{p_i,p_{i+1}} \right\}.\label{eq:shortpatht}
	\end{align}
	Now in this form this can be solved using the vast literature on shortest path algorithms.}\\

\section{Shortest Path Algorithms}

Shortest path algorithms~\citep{Gallo1988, dijkstra1959note}, such as Dijksta's algorithm, are popular algorithms which find the so called {\textit shortest path} within a graph. In our case the graph is formed such that the vertices or nodes uniquely correspond to a grid system index, i.e. a row/column in the transition matrix $T$. If there is a non-zero probability in $T_{i,j}$ we add an edge denoted $e_{i,j}$, where the weight on this edge is denoted $w(e_{i,j})=-\log(T_{i,j})$ between the vertex $i$ and going to the vertex $j$. Note that $T_{i,j}$ is not necessarily the same as $T_{j,i}$, hence we have a {\em directed graph}. Given a start vertex $o$ and an end vertex $d$, shortest path algorithms will find the path $P=\{v_1 \dots, v_n\}$ such that $P$ minimises the following
\[
\sum_{i=1}^{n-1} w(e_{v_i,v_{i+1}}),
\]
hence it solves the problem in \Cref{eq:shortpatht}. The algorithm used is exact, hence if no path is found then no path exists given the current network.
\subsection{Derivation of Equation \ref{eq:negbinom}}
\label{ssec:apx_deriv}
The derivation uses the Markov property, the conditional probability definition, and the fact that $P(x \in \{a,b\})=P(x=a)+P(x=b)$.
\begin{align}
&P(s_{t+k}=p_{i+1}, \{s_{t+l}=p_i\}_{l=1}^{k-1}\given s_{t}=p_{i}, \mathbf{p}\})\nonumber\\
	&=P(s_{t+k}=p_{i+1}\given s_{t+k-1}=p_i, s_{t+k}\in \{p_i,p_{i+1}\})
	\nonumber	\\	&\quad \times \prod_{l=1}^{k-1} P(s_{t+l}=p_i \given s_{t+l-1}=p_i, s_{t+l}\in \{p_i,p_{i+1}\})\nonumber\\
	&=\dfrac{P(s_{t+k}=p_{i+1}\given s_{t+k-1}=p_i)}{P(s_{t+k}\in \{p_i,p_{i+1}\}\given s_{t+k-1} = p_{i})} \nonumber\\  &\quad \times \prod_{l=1}^{k-1} \dfrac{P(s_{t+l}=p_i \given s_{t+l-1}=p_i)}{P(s_{t+l}\in \{p_i,p_{i+1}\}\given s_{t+l-1}=p_i)}\nonumber\\
	&= \dfrac{P(s_{t+k}=p_{i+1}\given s_{t+k-1}=p_i)}{P(s_{t+1}\in \{p_i,p_{i+1}\}\given s_{t}=p_i)^k}\nonumber\\ 
	&\quad \times \prod_{l=1}^{k-1} P(s_{t+l}=p_i \given s_{t+l-1}=p_i)\nonumber\\
	&= \dfrac{T_{p_i,p_{i+1}} T_{p_i,p_i}^{k-1} }{(T_{p_i,p_i}+T_{p_i,p_{i+1}})^{k}}\nonumber
\end{align}
where the first equality follows from the explanation given in Section \ref{ssec:ttest}.
\section{Brief Sensitivity Analysis to cut off time}
\label{ssec:apx_sens}
\begin{figure}[ht]
	\centering
	\includegraphics[width=228pt]{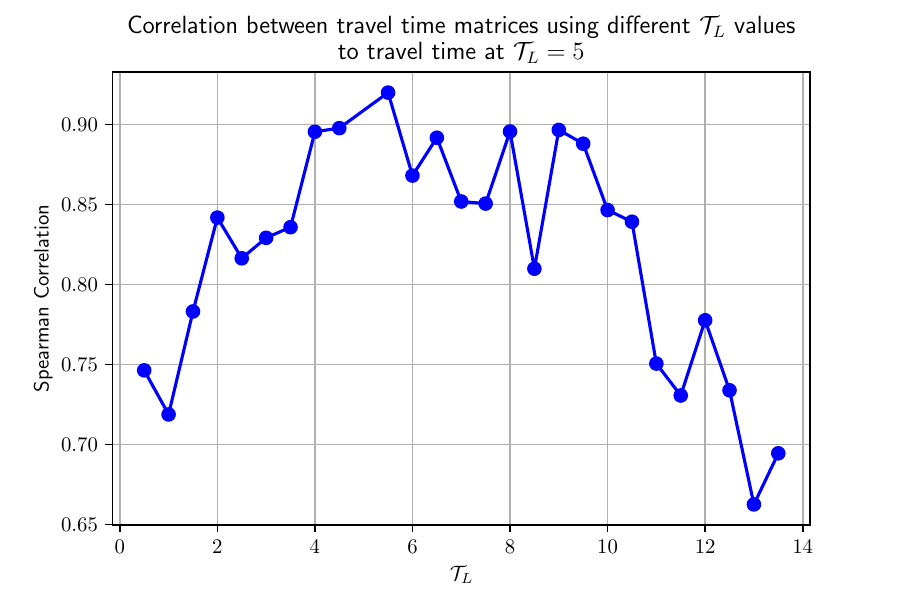}
	\caption{Spearman Correlation coefficient between the non-diagonal elements of the travel time matrix generated by $\mathcal{T}_L=5$ and the matrices generated by the values of $\mathcal{T}_L$ on the $x$-axis.}
	\label{fig:correlation}
\end{figure}

The main tuning parameter which we have fixed in this paper is the Lagrangian cut off time used when estimating the transition matrix $T$. The method is not especially sensitive to this choice as we shall now demonstrate. To show the sensitivity we ran an experiment where for a grid of values for $\mathcal{T}_L$ we estimated a pairwise travel time matrix for the locations in \Cref{tbl:locations}, then we estimated the Spearman correlation coefficient between the non-diagonal entries of each matrix to the corresponding entry of the travel time matrix generated from $\mathcal{T}_L=5$. Results are shown in \Cref{fig:correlation}. The experiment shows that the distances change but overall the matrices are very strongly correlated, particularly for $\mathcal{T}_L>2$. For comparison the average correlation value between the the pairwise travel time matrix $\mathcal{T}_L$ and the travel time matrices generated from the 100 rotations used in Section \ref{ssec:res_rot} is 0.8. A similar analysis, considering sensitivity to grid sizes is given in the online supplementary information.\\

\bibliographystyle{apalike}

\end{document}


\newpage
\maketitle
	\abstract{
     This document contains supplementary information including extra examples for the interested reader.
     We include: 
     \begin{enumerate}
         \item An example of various grid systems to produce pathways. This allow the reader to see the effects different grid systems have.
         \item A brief comparison to an example given by Smith et al. (2018).
         \item Three additional versions of Figure 9 for comparison purposes.
         \item A global shortest travel time map, in contrast to the most likely path travel time map given in Figure 7 of the main text.
         \item A more in depth sensitivity analysis.
         \item An explanation of how the method can find artificial connections and an adaptation of the method in the Strait of Gibraltar case.
     \end{enumerate}
	}
\section{Square Grid Paths vs H3}
\begin{figure}
        \centering
        \includegraphics[width=\linewidth]{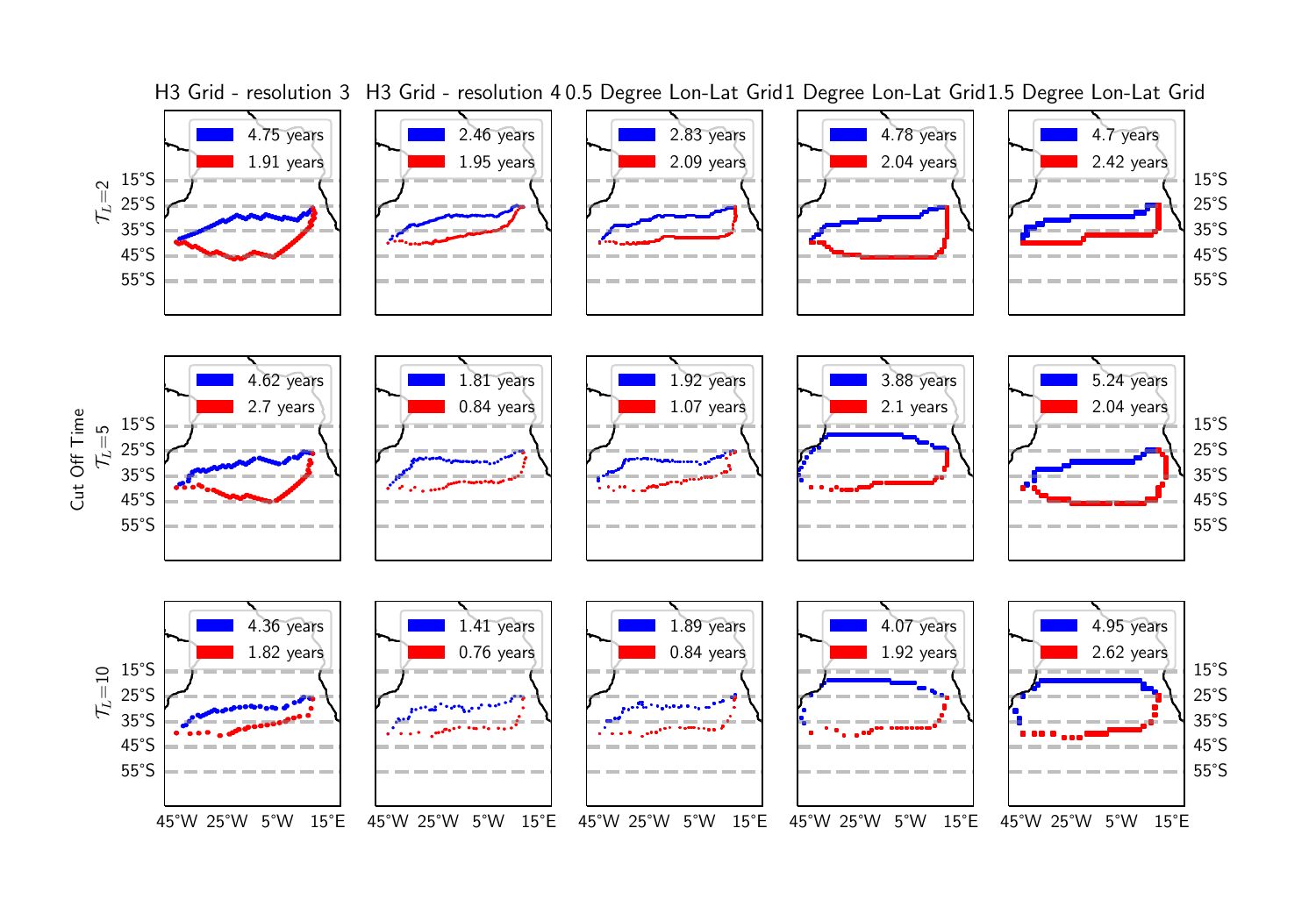}
        \caption{Paths and travel times between location 1 and 3 in Table 1 of the main text. Results given for $\mathcal{T}_l$ = 2, 5, 10 days and with various grid systems. Grid systems are the same within a column, indicated by the title of that column. Lagrangian cut-off times are altered by row. }
        \label{fig:tchange}
    \end{figure}

    We ran an experiment using a \textbf{resolution 3} and \textbf{resolution 4} H3 grid and three different sized longitude-latitude grids. The resolution 4 grid breaks down each resolution 3 cell into roughly seven polygons (in the manner seen in Figure 3 of the main text), resulting in an average area of $1,770km^2$, hence a much finer grid. The transition probabilities from the smaller grid sizes generally have higher uncertainty, due to less data which then results in even more variable pathways. In Figure \ref{fig:tchange}, we can see that the variance across grid sizes is far larger than the difference in results from changing $\mathcal{T}_L$. 
    
    One of the more concerning features of a longitude-latitude grid system is that the pathways tend to choose transitions either directly east, west, north or south and rarely do diagonal transitions. This is particularly visible with the larger 1.5 degree grid size. In contrast, for the H3 grids we do not see this sort of pattern, and trajectories are more naturally resembling meandering pathways. This is a key benefit of the H3 grid system, in addition to the more constant cell sizes it produces across the globe. 
\section{Comparison to \cite{Smith2018}}
\begin{figure}[h]
    \centering
    \includegraphics{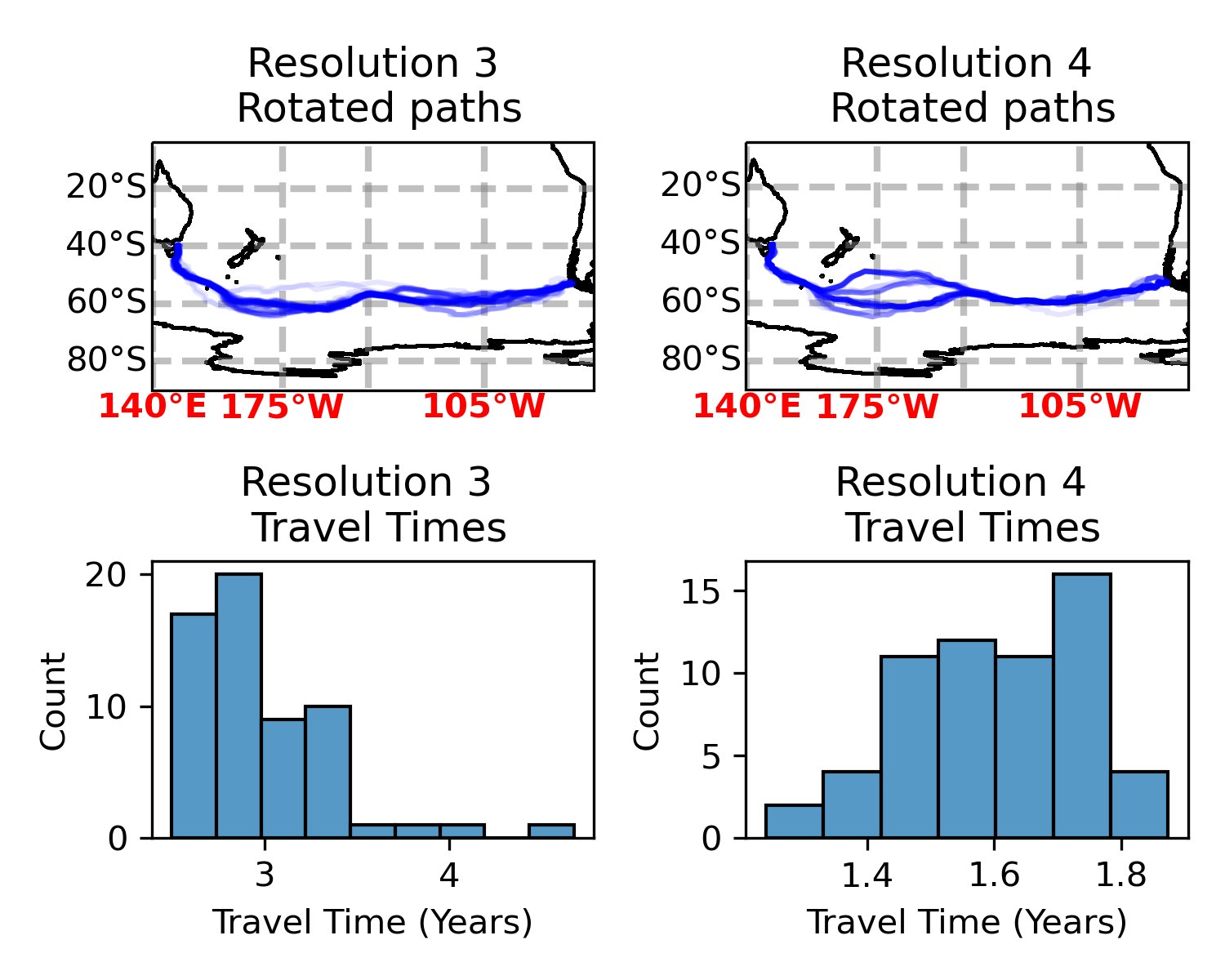}
    \caption{60 Pathways of particles going from the south-east coast of Australia to the south-west coast of Brazil. Both resolution 3 and resolution 4 H3 grid results shown. Pathway axis and grid lines set to match Figure 2 of \cite{Smith2018}.}
    \label{fig:rare_pathways}
\end{figure}

\cite{Smith2018} studied long-distance dispersal events through a number of methods. An example of pathways are given in Figure 2 of \cite{Smith2018}, about how an object could drift from the south-east coast of Australia over to the south-west coast of Brazil. The two pathways given are from simulated particles under the HYbrid Coordinate Ocean Model (HYCOM)~\citep{chassignet2007hycom} and using Monte Carlo Super Trajectories \citep{Sebille2011}, which we qualitatively compare to our Most Likely Path method here. Figure \ref{fig:rare_pathways} provides the results which the most likely path method found using 60 rotations. We show results from both resolution 3 and resolution 4 in the figure. 

The pathways shown in Figure \ref{fig:rare_pathways} look very similar to those of the route of simulated HYCOM particles. The travel times reported in Figure \ref{fig:rare_pathways} are also comparable, where the minimum time reported in \cite{Smith2018} is 2.4 years using MCSTs. The resolution 3 travel time results in a close match, which is expected as the grid size is similar to what Monte Carlo Super Trajectories are created with.

\section{Bootstrap and Rotation Pathways}
In Figure 9 of the main body we showed example pathways for rotations in resolution 3 of the H3 grid system. Here we show the same results in three different flavors.

\begin{figure}
    \centering
    \includegraphics[width=\textwidth]{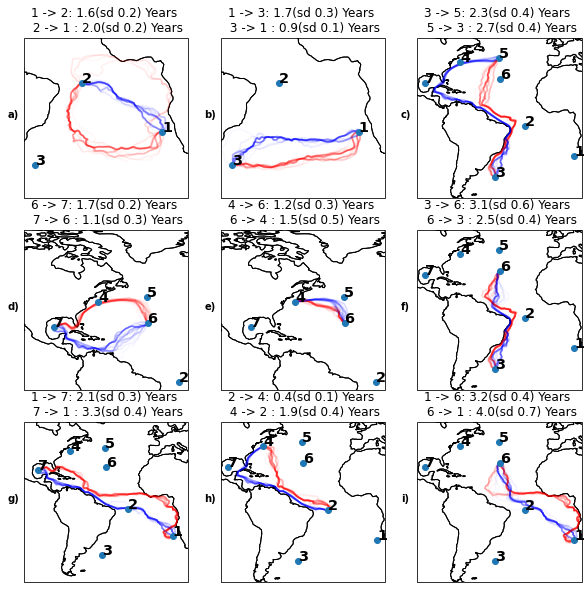}
    \caption{This figure is similar to Figure 9 of the main text, however here we only use 60 rotations and use resolution 4 of the \textit{H3} grid system. Each line connects the centroid of each hexagon within the path.}
    \label{fig:res4_rot}
\end{figure}

\begin{figure}
    \centering
    \includegraphics[width=\textwidth]{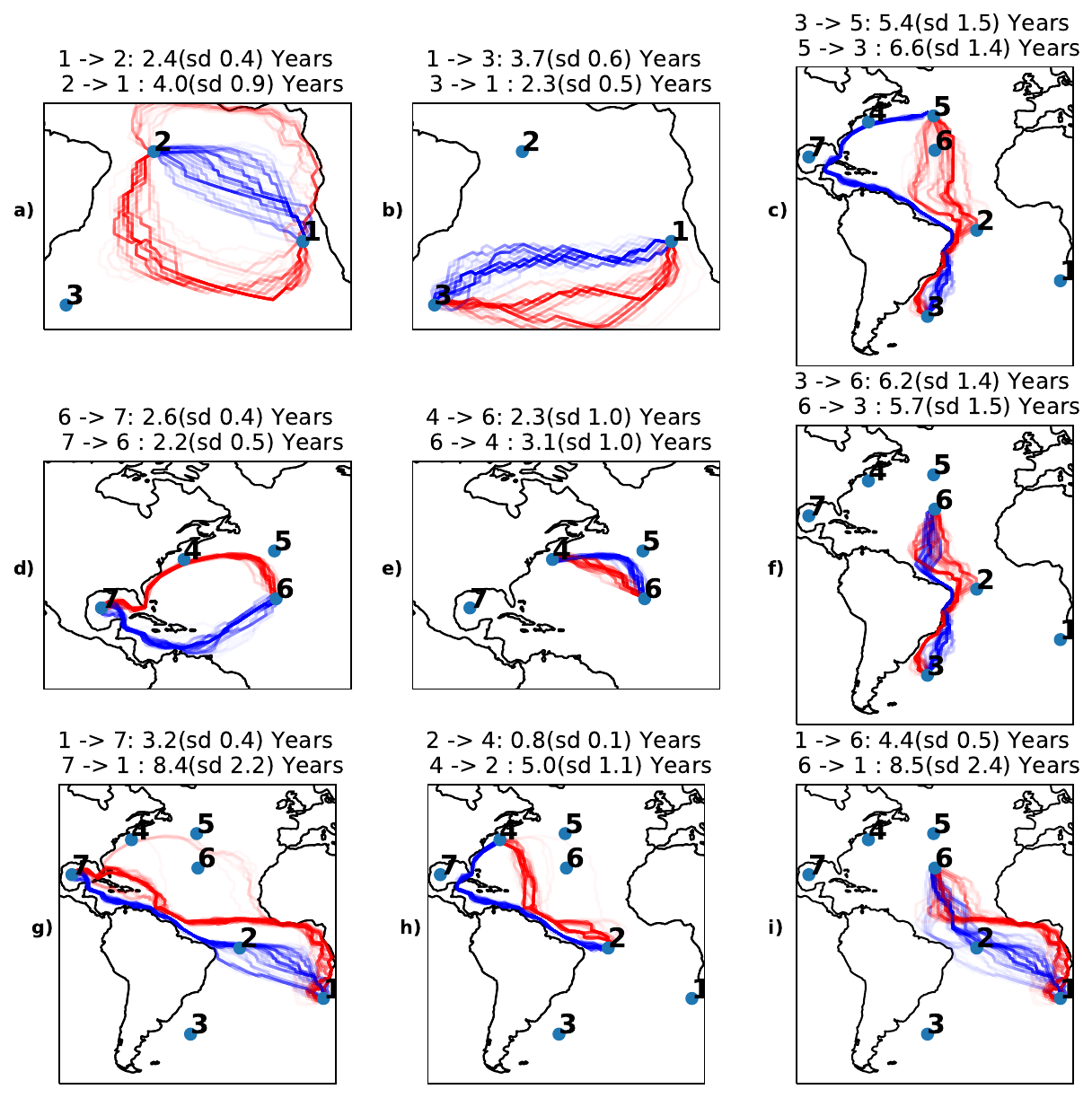}
    \caption{This figure is similar to Figure 9 of the main text. Here we only use 100 bootstrap samples instead of rotations and use resolution 3 of the \textit{H3} grid system. Each line connects the centroid of each hexagon within the path. Note all paths are in the same non-rotated grid system here.}
    \label{fig:res3_boot}
\end{figure}

\begin{figure}
    \centering
    \includegraphics[width=\textwidth]{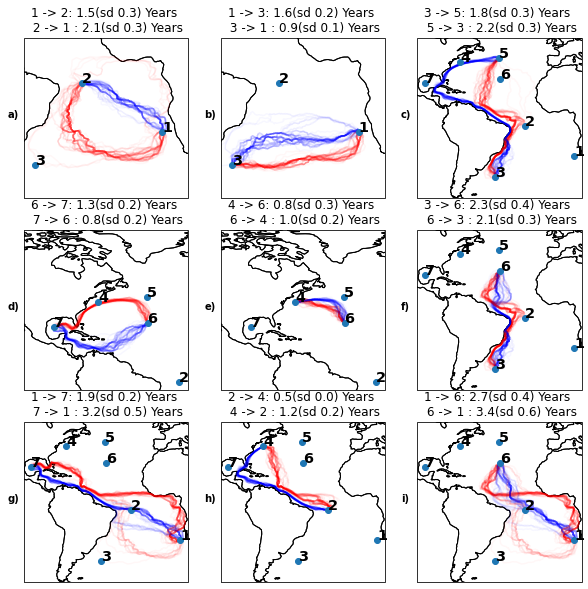}
    \caption{This figure is similar to Figure 9 of the main text. Here we only use 100 bootstrap samples instead of rotations and use resolution 4 of the \textit{H3} grid system. Each line connects the centroid of each hexagon within the path. Note all paths are in the same non-rotated grid system here.} 
    \label{fig:res4_boot}
\end{figure}

\begin{itemize}
    \item A similar plot using resolution 4 of the H3 grid system in Figure \ref{fig:res4_rot} with 60 rotations. 
    \item A bootstrap version in Figure \ref{fig:res3_boot} with 100 bootstrap samples.
    \item A bootstrap version using resolution 4 in Figure \ref{fig:res4_boot} with 100 bootstrap samples.
\end{itemize}

Due to computational restrictions with the resolution 4 networks we use 60 rotations instead of the 100 used in the main document for resolution 3.


We see that there is less variance in the pathways due to rotations in resolution 4 compared to resolution 3, however the variability in the pathways from the bootstrap is considerably larger. This is due to the classic bias-variance trade-off in statistical estimation. The results are less biased at resolution 4 due to less discretization, but the variance of the transition matrix entries increases due to less data being available to estimate transition probabilities. In the main body we present resolution 3 as this seems to balance this trade-off well in the global dataset, however in regional studies with high data density (such as data from numerous recent clustered drifter deployments), we imagine that resolution 4 might optimally balance this trade-off and reduce uncertainty in these regional studies.

\section{Shortest Travel Time Map}
\begin{figure}
\centering
\includegraphics[height=0.7\textheight]{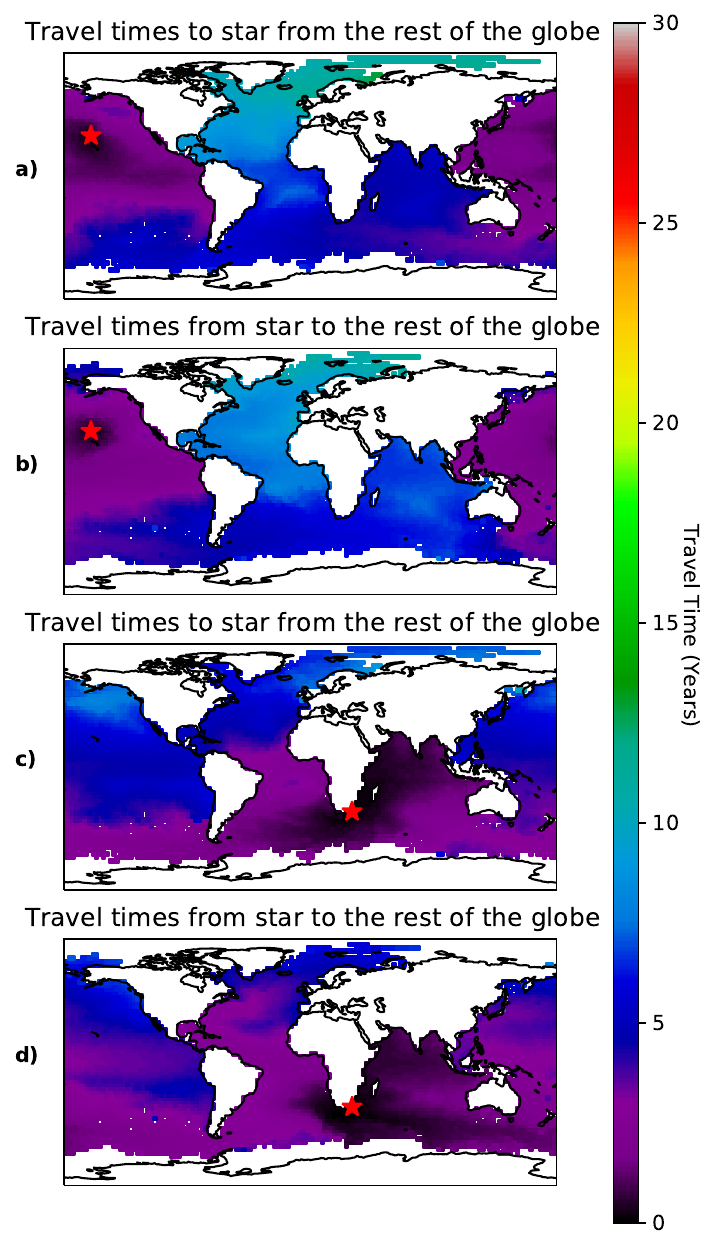}
\caption{Similar to Figure 5 of the main text, however rather than the travel time of the most likely path, this shows the shortest expected travel time.}
\label{fig:shortest}
\end{figure}

We recreate the travel time map shown in Figure 5 of the main document, however this time we use the expected travel time as the objective of the shortest path algorithm. The new objective we use to find the optimal path and travel time is:
\begin{equation}
    \hat{p}=\argmin_{\mathbf{p}\in \mathcal{P}} \sum_{i=0}^{n-1} \dfrac{T_{p_i,p_i}}{T_{p_i,p_{i+1}} +1},
\end{equation}
which is the minimum of Equation (9) in the main text. 
We show the result of this in Figure \ref{fig:shortest}. As we are now directly looking for a minimum, this map is more directly comparable to the results shown in Figure 2 in \cite{jonsson2016timescales}. The map shown in Figure \ref{fig:shortest} is smoother than the results of \cite{jonsson2016timescales}. 

The travel times from this map may be preferable to the most likely travel times map; for example if distances which obey the triangle inequality are desired. These are not used as part of the main text as we believe the expected travel time of the most likely path is a more suitable measure for most applications. Essentially, if there was a `true' and non-Gaussian travel time distribution, the minimum travel time is an estimate of the minimum of that distribution, whereas the travel time of the most likely path is more akin to a mode. This explains, in part, the larger travel times we obtain in Figure 5 of the main document, versus what we see here in Figure \ref{fig:shortest} and in Figure 2 of \cite{jonsson2016timescales}.

\section{Grid Size and Lagrangian cut off time sensitivity}
\begin{figure}
    \centering
    \includegraphics[width=0.5\textwidth]{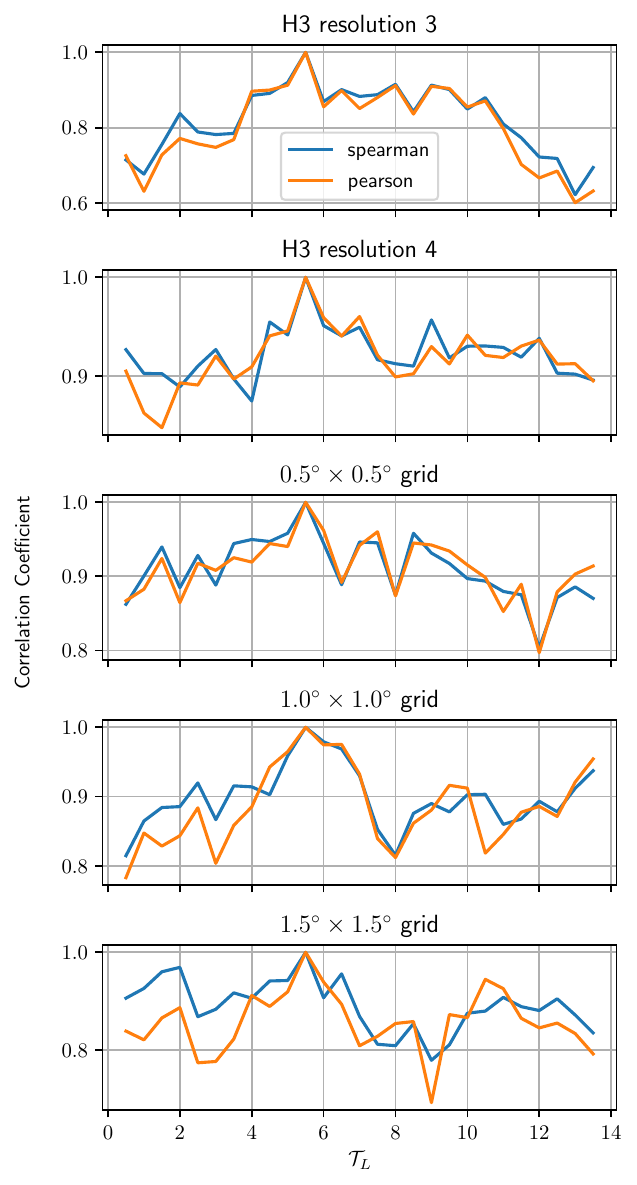}
    \caption{A repeat of the experiment shown in Figure 10 of the main text, however we show the results under both Spearman and Pearson correlations and under 5 different grid systems.}
    \label{fig:correlation_exp}
\end{figure}

To investigate the relationship between the assumed decorrelation time ($\mathcal{T}_L$), and the grid size, we show an extended version of the sensitivity analysis shown in Appendix d of the main paper. Under each resolution we estimate the travel time matrix for a grid of values for $\mathcal{T}_L$. Then we estimate the Spearman and Pearson correlation (of the off-diagonal entries) to the travel times created at $\mathcal{T}_L=5$ for that grid system. The results are shown in Figure \ref{fig:correlation_exp}.
\begin{figure}
    \centering
    \includegraphics[width=\textwidth]{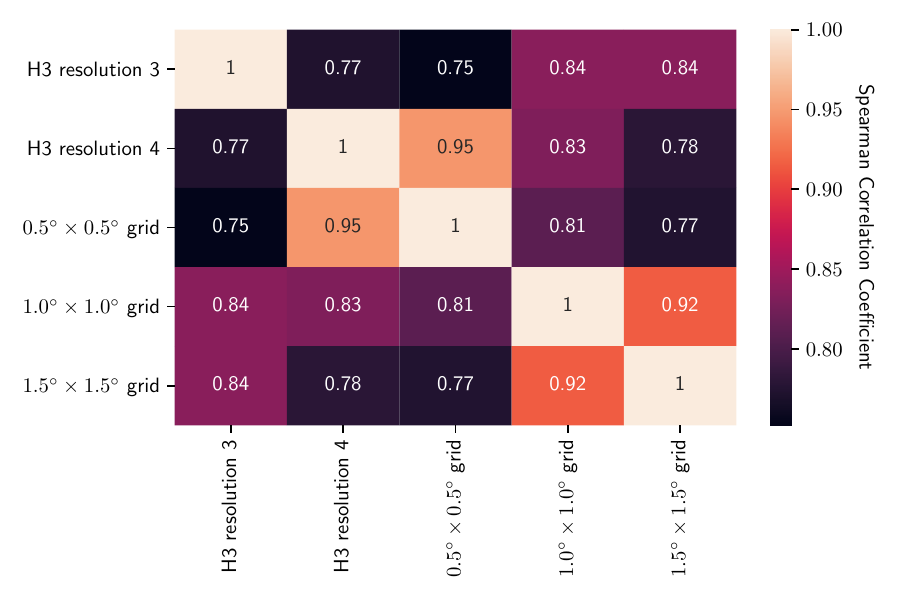}
    \caption{The pairwise Spearman correlation value between the 42 travel time estimates produced by each grid system. We fix $\mathcal{T}_L=5$ days.}
    \label{fig:corrheatmap}
\end{figure}
We show the same correlation metric between each pair of grid systems in Figure \ref{fig:corrheatmap}. We see the correlation values are at worst 0.68 which would still be interpreted as a moderate to strong positive correlation. 

\begin{figure}
    \centering
    \includegraphics{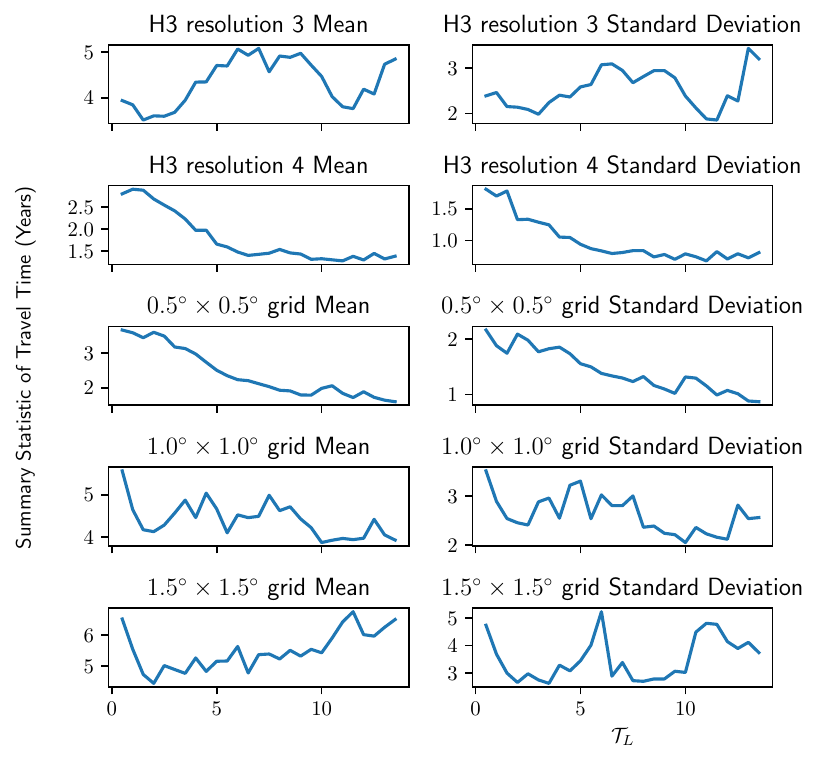}
    \caption{The mean and standard deviation of the 42 travel time estimates used throughout the paper. We show the mean and variance of these 42 travel time estimates in years for each combination of the 5 grid systems and 28 decorrelation times (0.5, 1, 1.5,..., 14 days).}
    \label{fig:meanvar}
\end{figure}
We also show the mean and variance of the travel times in each matrix produced under the five grid systems and a grid of values for $\mathcal{T}_L$ in Figure \ref{fig:meanvar}. It can clearly be seen that the smaller grid systems (\textit{H3} resolution 4 and $0.5^\circ \times 0.5^\circ$) are less robust to changes in value of $\mathcal{T}_L$. The mean and standard deviation show a downward trend as we increase the value of $\mathcal{T}_L$. This shows the added robustness of resolution 3 and motivates this choice in the main paper for this dataset (the Global Drifter Program). We recommend repeating such sensitivity analyses for use with different datasets, especially if applied to simulated trajectories or regional studies.

\section{Artificial connections}
	When running the analysis for the rotations of Section 5c, if we do not take the preprocessing step of removing the two points on the Strait of Gibraltar, we find that some rotations allow this connection. In 71 of the 100 rotations we were unable to obtain a travel time estimate from the Atlantic into the Mediterranean and in 95 we were unable to find a travel time estimate from the Mediterranean to the Atlantic. When we do not do a rotation we are able to obtain an estimate into the Mediterranean, this is due to the way the grid aligns as shown in Figure 3. Even if only one of the 100 rotations are unable to provide an estimate it would be advisable to not use the estimate from this method. Therefore, using the vanilla method on its own to estimate travel times into the Mediterranean is not a good option.
	
	Overall, the method provided depends on the availability of drifter data making a connection at some point. Connections such as going across the Strait of Gibraltar are in practice highly unlikely; any pathway which crosses it is due to a grid covering both the east and west of the Strait of Gibraltar. One potential way to adapt the method to approximate travel times across the Strait is, either adding artificial simulated trajectories as in \citet{VanSebille2012}, or simply add a very small probability to the transition matrix crossing from the west to the east of the Strait of Gibraltar (and vice versa). For example, take two locations, one west and one east of the Strait of Gibraltar, say these correspond to states $w$ and $e$ respectively. If we wanted the crossing time to be 100 days into the Mediterranean sea, set $T_{e,w}$ such that $19\times T_{e,w}= T_{e,e}$, the transition matrix will no longer be valid as the $e$ row no longer sums to one but the method will still work as intended, giving a 100 day crossing time from state $e$ to $w$. Such an adaptation would require the removal of the state which covers the Strait of Gibraltar to force the algorithm to take the artificial 100 day crossing.
	
    This example where the method detects the Mediterranean Sea's artificial connection is an interesting bonus feature of the rotation methodology, however, it is not as easily applicable to the Panama land mass problem. In the case of Panama, we will still obtain a travel time estimate from the Gulf of Mexico to the Pacific if a grid cell can cover both sides, but the times which are permitted to skip over the Panama land mass will be much shorter. An automatic detection could be achieved by looking at a large sample of rotations then running a test for multi modality. If it finds that there are two modes which are very far apart then this would be a sign that the method is finding some shortcut which is only present under some rotations. If such a method worked to detect the Panama land mass, we could then use it to search for more subtle surface transport barriers. In general it is preferable to preprocess the transition matrix $T$ such that rows/columns corresponding to unwanted links such at the Panama Canal and the Strait of Gibraltar are simply removed, as we performed in our analysis. Visual detection of pathways will generally solve any issues.

\bibliographystyle{apalike}
\bibliography{supplement}